\documentclass[a4paper,11pt]{article}
\pdfoutput=1

\usepackage{jheppub}
\usepackage[T1]{fontenc}
\usepackage{lmodern}
\usepackage{CJKutf8}
\usepackage{amsmath,amssymb,mathtools,mathrsfs}
\usepackage{booktabs,microtype}
\usepackage{flafter}
\usepackage{tikz}
\usetikzlibrary{arrows.meta,positioning,calc}
\definecolor{periodblue}{RGB}{35,85,132}
\definecolor{insertionred}{RGB}{170,62,40}

\numberwithin{equation}{section}
\newcommand{\dd}{\mathrm d}
\newcommand{\dlog}{\dd\log}
\newcommand{\qb}{\bar q}
\newcommand{\Qb}{\bar Q}
\newcommand{\cI}{\mathcal I}
\newcommand{\cR}{\mathcal R}
\newcommand{\bR}{\mathbf R}
\newcommand{\cK}{\mathcal K}

\newcommand{\br}[1]{\langle #1\rangle}
\newcommand{\Res}{\operatorname*{Res}}
\newcommand{\Reg}{\operatorname{Reg}}
\newcommand{\Li}{\operatorname{Li}}

\hypersetup{
  pdftitle={A periodic q-bar equation for form factors},
  pdfauthor={Song He and Jiahao Liu}
}

\title{\boldmath A periodic $\bar q$ equation for form factors}

\author[a,b,c]{Song He (何颂)}
\author[a,c]{Jiahao Liu (刘家昊)}

\affiliation[a]{New Cornerstone Laboratory, Institute of Theoretical Physics,
Chinese Academy of Sciences, Beijing 100190, China}
\affiliation[b]{School of Fundamental Physics and Mathematical Sciences,
Hangzhou Institute for Advanced Study, UCAS \& ICTP-AP,
Hangzhou 310024, China}
\affiliation[c]{School of Physical Sciences, University of Chinese Academy of
Sciences, No.~19A Yuquan Road, Beijing 100049, China}

\emailAdd{songhe@itp.ac.cn}
\emailAdd{liujiahao@itp.ac.cn}

\abstract{
MHV form factors of the chiral stress-tensor multiplet in planar $\mathcal N=4$ super Yang–Mills theory are dual to periodic null Wilson loops. We propose a periodic $\bar q$ equation for the MHV and higher-helicity sectors at vanishing fermionic monodromy. The equation relates kinematic derivatives of $\mathrm N^k\mathrm{MHV}$ form factors to collinear integrals of $\mathrm N^{k+1}\mathrm{MHV}$ form factors with one additional external particle, providing a recursive relation in loop order. In periodic momentum supertwistors, it has the same collinear-integral form as the two dotted components of the amplitude $\bar Q$ equation, with coefficient $\Gamma_{\rm cusp}/4$, despite the absence of ordinary dual conformal invariance at fixed operator momentum. For parity-even scalar MHV remainders, combining the equation with its parity image supplies all directions of the kinematic differential. 

We verify the equation analytically for the one-loop four-point NMHV form factor and reconstruct the two-loop three-point MHV remainder at function level. Further two-loop MHV tests give numerical agreement with derivatives of the known four-point remainder at a fixed kinematic configuration, and for five points, all six chiral derivatives agree exactly at symbol level for a one-parameter family of kinematics. The structure of the collinear source further gives all-loop predictions for MHV last entries, consistent with known results and form a space of dimension $n(2n-5)$. The periodic Wilson-loop picture also motivates an all-loop derivation. Using the form-factor OPE, we extract the one-loop correction to the fermion energy from the logarithmic source and recover its vanishing at zero momentum. 
}


\begin{document}
\begin{CJK*}{UTF8}{gbsn}
\maketitle
\end{CJK*}
\raggedbottom

\section{Introduction}
\label{sec:introduction}

Form factors are matrix elements of local gauge-invariant operators between
the vacuum and on-shell multiparticle states. For the chiral stress-tensor
multiplet in planar $\mathcal N=4$ super Yang--Mills (SYM) theory, their
on-shell superspace and helicity expansion closely parallel those of
scattering amplitudes~\cite{Brandhuber:2011tv,Bork:2011mhv,Bork:2012tt,Bork:2014eqa}.
The momentum $q$ carried by the operator changes the dual geometry:
the null contour formed by the external momenta closes only up to a
translation by $q$. MHV form factors are dual to periodic null Wilson
loops~\cite{Brandhuber:2010ad}. This duality
makes the periodic contour the natural setting in which to seek a
supersymmetric relation between form factors of different multiplicities
and helicities.

The periodic Wilson-loop description has developed into a practical tool
for studying these observables. Its extension to superspace gives a
super Wilson-loop duality for half-BPS operators~\cite{Basso:2023bwv},
and the bosonic construction has yielded an all-multiplicity two-loop MHV
integrand and explicit integrated five- and six-point
results~\cite{Li:2024rkq}. The form-factor operator product expansion
(FFOPE) resolves the periodic polygon into propagating flux-tube
excitations and a terminal transition coupling them to the
operator~\cite{Sever:2020jjx,Sever:2021born,Sever:2021xga}. It supplies near-collinear
data for the three-point bootstrap, which has reached eight
loops~\cite{Dixon:2020bbt,Dixon:2022rse}. At four points, the two-loop MHV
form factor is known as a complete analytic function
\cite{Guo:2024bsd,Dixon:2024yvq}, and its three-loop symbol is also
known~\cite{Dixon:2022xqh}. The four-point NMHV form factor is known
through three loops at symbol level~\cite{He:2026ddp}. Together with
the study of form-factor symbol alphabets~\cite{He:2026kdq}, these results
make it possible to test relations that connect different loop orders,
multiplicities and helicity sectors.

For amplitudes, such a relation is provided by the $\bar Q$ equation
\cite{Caron-Huot:2011zgw,Bullimore:2011kg,Caron-Huot:2011dec}. In momentum
supertwistors~\cite{Hodges:2009hk}, the generator takes the form
\begin{equation}
 \Qb_a^A=\sum_i\chi_i^A\frac{\partial}{\partial Z_i^a} .
 \label{eq:intro-qbar}
\end{equation}
Its anomalous action on a BDS-normalized $\mathrm N^k\mathrm{MHV}$ amplitude
is a subtracted collinear integral of an amplitude with one additional
particle in the $\mathrm N^{k+1}\mathrm{MHV}$ sector. The coefficient is
$\Gamma_{\rm cusp}/4$, so at loop order $L$ the source involves amplitudes
through at most $L-1$ loops. The supersymmetric Wilson loop
\cite{Mason:2010yk,Caron-Huot:2010ryg} explains the local origin of this
descent relation: the difference between the full and self-dual
supersymmetry transformations produces a field insertion that also arises
from a supersymmetric collinear kink. In the flux-tube description, the
supersymmetry charge is carried by a zero-momentum fermion
\cite{Caron-Huot:2011dec,Basso:2014hfa}.

The amplitude $\bar Q$ equation has been used directly to determine
multi-loop symbols and their algebraic letters
\cite{He:2019jee,He:2020vob,Li:2021bwg}. It has also played a crucial role
in the higher-loop bootstrap program by providing last-entry conditions
and relations among their coefficients
\cite{Caron-Huot:2011dec,CaronHuot:2019SixSevenLoops}.
The connection follows from the differential form of the equation:
the $\dd\log$ factors multiplying lower-weight functions determine the
last entries of the amplitude symbol. Restrictions on these differentials
therefore constrain a bootstrap ansatz before the collinear integrals
are evaluated in full. Beyond MHV, the allowed last entries are tied to
the rational superinvariants multiplying the transcendental functions.
The resulting relations, together with symmetry and boundary data,
substantially reduce the freedom in high-loop amplitudes.

For form factors, the period $q$ is a physical external datum. Periodic
momentum twistors provide a convenient description of the contour
\cite{Bianchi:2018rrj}, but the fixed-$q$ kinematics do not have the
ordinary dual conformal invariance of the closed amplitude polygon. The full
twistor generator \eqref{eq:intro-qbar} likewise does not preserve the
chosen fixed-period kinematics. At vanishing fermionic monodromy
($\gamma=0$), its two dotted components do preserve this slice. We denote
this restricted generator by $\qb$, while $q$ denotes the momentum
carried by the operator. We propose a periodic collinear integral
equation with the same form as these two components of the amplitude
$\bar Q$ equation. The insertion adds a cusp together with its complete
periodic orbit, and the source and subtraction are regulated with a
common endpoint prescription. The proposed equation retains the
coefficient $\Gamma_{\rm cusp}/4$ and uses the form-factor BDS subtraction.

This equation would connect derivatives of higher-loop form factors to
lower-loop data in the next helicity sector. It also provides a way to
investigate the last-entry conditions and coefficient relations relevant
to form-factor bootstraps. In particular, it relates the three-point MHV
remainder to a collinear integral of the four-point NMHV form factor,
linking two sectors for which substantial bootstrap data are already
available. The corresponding differential constraints would complement
the near-collinear expansion supplied by the FFOPE. For NMHV form factors,
extracting such constraints requires keeping track of the rational
prefactors as well as the transcendental functions; a last-entry alphabet
alone does not capture all the information in the equation.

The chiral nature of $\qb$ requires a further step when reconstructing an
MHV remainder. At generic kinematics and fixed $q$, the equation supplies
$2n-4$ independent derivatives, while the physical kinematic space has
dimension $3n-7$ after the little-group and fixed-$q$ Lorentz quotients.
Thus the three-point differential is already complete, whereas higher
multiplicity leaves $n-3$ directions undetermined. For a parity-even
scalar MHV remainder, evaluating the same source at the parity related
configuration supplies the missing directions. We give an explicit
construction of the full differential, including all five coordinate
derivatives at four points, and specify the consistency conditions needed
to integrate it to the remainder. Boundary data fix the remaining
additive constant.

We test the equation in both the one-loop NMHV and two-loop MHV sectors.
The four-point one-loop NMHV variation agrees analytically with a
five-point tree-level source. For the three-point MHV form factor,
the four-point one-loop NMHV source yields a closed differential that
integrates to the complete two-loop remainder
\cite{Brandhuber:2012vm,Dixon:2020bbt}, including the terms beyond its
symbol. At four points, we numerically compare the two-loop MHV derivatives
with those of the known remainder at a generic configuration. At five points all six
independent chiral derivatives agree with the known two-loop symbol on a
one-parameter family after projecting out symbol letters constant along
that family. We also implement the parity reconstruction
numerically at four and five points. These comparisons test both the
collinear source and its use in recovering physical derivatives.

We also study MHV last entries through the collinear contributions of
five-brackets, five-brackets with rational four-bracket multipliers, and
three-mass triangle leading singularities. Combining the two triangle
cut solutions gives rational $\qb\log$ factors independent of the
integration variable. The resulting rational last-entry candidates have
the proposed counting $n(2n-5)$, with the assumptions underlying this
count specified in section~\ref{sec:mhv-last-entries}.
The three-point candidates reproduce the known last-entry condition.
At four points, the nine-dimensional last-entry space of the two-loop
remainder lies within the twelve-dimensional candidate space, while the
three-loop space coincides with it~\cite{Dixon:2022xqh}.
For $n=3,\ldots,6$, the candidates also span the same space as the
last-entry structures of ref.~\cite{Li:2024rkq}.

Within the periodic Wilson-loop description, we also examine the
ingredients needed for an all-loop derivation. The local supersymmetry
anomaly and collinear-kink construction identify the field insertion
that the periodic equation should describe. For the three-point source,
we derive an analytic relation between the Mellin transform of its bare
tree overlap and that of the one-loop collinear logarithm. The multiplier
is the one-loop fermion dispersion relation, with a vanishing analytic
zero mode. This identifies the logarithmic propagation probed by the
FFOPE. Establishing the equation at finite coupling still requires the
finite normalization relating the Ward and kink insertions, including
the terminal transition, contact terms and the endpoint prescription on
the periodic contour.

Section~\ref{sec:equation} defines the equation and its insertion
prescription. Section~\ref{sec:differential} derives the parity completion
of the MHV differential, and section~\ref{sec:evidence} presents the
perturbative evidence and MHV last-entry analysis.
Section~\ref{sec:all-loop} studies the local
Wilson-loop insertion and the fermion kernel as preparation for an
all-loop derivation. We discuss further applications in
section~\ref{sec:discussion}. Appendix~\ref{app:monodromy} collects our
conventions. Appendices~\ref{app:comparison-details},
\ref{app:mhv-last-entries} and~\ref{app:fermion-mellin} provide details of
the perturbative calculations, MHV last-entry analysis and Mellin
transform, respectively.

\section{The periodic \texorpdfstring{$\bar q$}{q-bar} equation}
\label{sec:equation}

\subsection{Periodic kinematics and normalization}

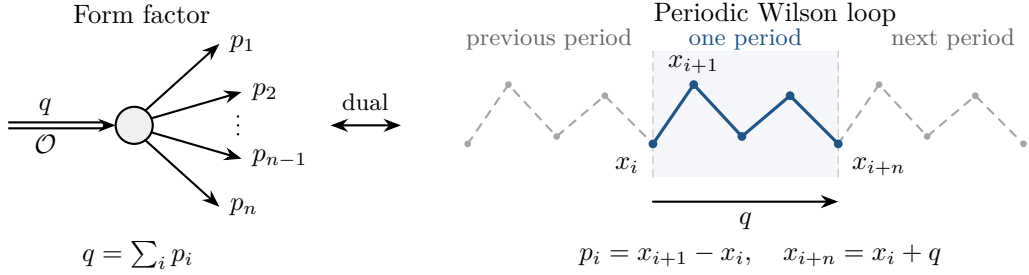
\begin{figure}[h]
\centering
\begin{tikzpicture}[x=0.98cm,y=0.98cm,>=Stealth,
  every node/.style={font=\small},line cap=round,line join=round]
  \node at (1.65,3.35) {Form factor};
  \draw[thick,line cap=butt] (-0.15,1.81)--(1.14,1.81);
  \draw[thick,line cap=butt] (-0.15,1.89)--(1.14,1.89);
  \draw[->,thick] (1.14,1.85)--(1.30,1.85);
  \path (-0.15,1.85)--(1.28,1.85)
    node[pos=.35,yshift=7pt] {$q$}
    node[pos=.35,yshift=-7pt] {$\mathcal O$};
  \filldraw[fill=black!6,thick] (1.55,1.85) circle (.25);
  \draw[->,thick] (1.70,2.05)--(2.7,2.95) node[right] {$p_1$};
  \draw[->,thick] (1.80,1.95)--(3.0,2.3) node[right] {$p_2$};
  \draw[->,thick] (1.80,1.75)--(3.0,1.4) node[right] {$p_{n-1}$};
  \draw[->,thick] (1.70,1.65)--(2.7,.75) node[right] {$p_n$};
  \foreach \y in {1.73,1.85,1.97}
    \fill (2.95,\y) circle (.45pt);
  \node at (1.6,.10) {$q=\sum_i p_i$};
  \draw[<->,thick] (4.20,1.85)--(5.15,1.85);
  \node[font=\footnotesize] at (4.67,2.17) {dual};

  \begin{scope}[shift={(1,0)}]
  \node at (9.20,3.35) {Periodic Wilson loop};
  \fill[periodblue!5] (7.55,1.15) rectangle (10.05,2.85);
  \draw[black!25,densely dashed] (7.55,1.15)--(7.55,2.85);
  \draw[black!25,densely dashed] (10.05,1.15)--(10.05,2.85);
  \foreach \dx in {5.05,10.05}{
    \begin{scope}[shift={(\dx,1.6)}]
      \draw[black!35,thick,dashed] (0,0)--(.55,.8)--(1.20,.10)--(1.85,.65)--(2.5,0);
      \foreach \x/\y in {0/0,.55/.8,1.2/.1,1.85/.65,2.5/0}
        \fill[black!35] (\x,\y) circle (1.25pt);
    \end{scope}
  }
  \begin{scope}[shift={(7.55,1.6)}]
    \draw[periodblue,very thick] (0,0)--(.55,.8)--(1.20,.10)--(1.85,.65)--(2.5,0);
    \foreach \x/\y in {0/0,.55/.8,1.2/.1,1.85/.65,2.5/0}
      \fill[periodblue] (\x,\y) circle (1.65pt);
    \node[below left,xshift=-3pt,yshift=-3pt,inner sep=2pt]
      at (0,0) {$x_i$};
    \node[above,inner sep=4pt] at (.55,.8) {$x_{i+1}$};
    \node[below right,xshift=3pt,yshift=-3pt,inner sep=2pt]
      at (2.5,0) {$x_{i+n}$};
  \end{scope}
  \node[font=\footnotesize,text=black!60] at (6.15,3.02) {previous period};
  \node[font=\footnotesize,text=black!60] at (11.6,3.02) {next period};
  \node[font=\footnotesize,text=periodblue] at (8.80,3.02) {one period};
  \draw[->,thick] (7.55,.82)--(10.05,.82) node[midway,below] {$q$};
  \node at (9.0,.10) {$p_i=x_{i+1}-x_i,\quad x_{i+n}=x_i+q$};
  \end{scope}
\end{tikzpicture}
\caption{Periodic null contour associated with an $n$-particle form factor.
The operator momentum $q=\sum_i p_i$ sets the period, $x_{i+n}=x_i+q$.
The highlighted segment represents one fundamental period, and the dashed
segments are its translated images.}
\label{fig:periodic-wilson-loop}
\end{figure}

We work on the periodic cover of an $n$-particle form factor, with
$i=1,\ldots,n$ labeling one fundamental period.  Each external state carries
the usual spinor-helicity variables
$(\lambda_i^\alpha,\widetilde\lambda_i^{\dot\alpha},\eta_i^A)$, with
$A=1,\ldots,4$.  We describe these data in dual superspace coordinates
$(x_i^{\alpha\dot\alpha},\theta_i^{\alpha A})$, defined by
\begin{equation}
 \begin{aligned}
 p_i^{\alpha\dot\alpha}
 &=x_{i+1}^{\alpha\dot\alpha}-x_i^{\alpha\dot\alpha}
 =\lambda_i^\alpha\widetilde\lambda_i^{\dot\alpha},
 \qquad \sum_{i=1}^n p_i^{\alpha\dot\alpha}=q^{\alpha\dot\alpha},\\
 x_{i+n}^{\alpha\dot\alpha}
 &=x_i^{\alpha\dot\alpha}+q^{\alpha\dot\alpha},
 \qquad
 \theta_{i+1}^{\alpha A}-\theta_i^{\alpha A}
 =\lambda_i^\alpha\eta_i^A .
 \end{aligned}
 \label{eq:periodic-x}
\end{equation}
The operator carries momentum $q$ and supermomentum
$\gamma^{\alpha A}$, so the fermionic dual coordinates obey
\begin{equation}
 \theta_{i+n}^{\alpha A}=\theta_i^{\alpha A}+\gamma^{\alpha A} .
 \label{eq:periodic-superspace}
\end{equation}
Thus, $q$ and $\gamma$ are respectively the bosonic and fermionic
monodromies of the periodic contour.  We use four-component fermionic
variables after stripping the universal supermomentum delta functions.
Throughout this paper we restrict to the zero-supermomentum slice
$\gamma=0$, or equivalently to vanishing fermionic monodromy.\footnote{In
the Grassmann Fourier representation, this selects the chiral on-shell
Lagrangian, the highest component of the chiral stress-tensor
multiplet~\cite{Brandhuber:2011tv}.}  We also assume generic kinematics
with $q^2\neq0$, so that the bispinor
$q^{\alpha\dot\alpha}$ is invertible in the parity construction of
section~\ref{sec:differential}.  The resulting periodic dual contour is
illustrated in figure~\ref{fig:periodic-wilson-loop}.

To describe the action of the dual superconformal generators, we pass to the
momentum supertwistors $\mathcal Z_i=(Z_i,\chi_i)$
\cite{Hodges:2009hk,Bork:2014eqa,Basso:2023bwv}, with
$Z_i=(\lambda_i,\mu_i)$,
$\mu_i^{\dot\alpha}=x_i^{\alpha\dot\alpha}\lambda_{i\alpha}$, and
$\chi_i^A=\theta_i^{\alpha A}\lambda_{i\alpha}$.  Their periodicity is
encoded by the period translation $P$.  On the $\gamma=0$ slice, it acts as
\begin{equation}
 \mathcal Z_{i+n}=P\mathcal Z_i,
 \qquad
 \lambda_{i+n}=\lambda_i,\qquad
 \mu_{i+n}^{\dot\alpha}
 =\mu_i^{\dot\alpha}+q^{\alpha\dot\alpha}\lambda_{i\alpha},
 \qquad
 \chi_{i+n}=\chi_i .
 \label{eq:periodic-twistors}
\end{equation}
Indeed, in general
$\chi_{i+n}^A-\chi_i^A=\gamma^{\alpha A}\lambda_{i\alpha}$; the last
equality in \eqref{eq:periodic-twistors} follows on the $\gamma=0$ slice.
The two dotted components of the amplitude $\bar Q$ generator
\cite{Caron-Huot:2011dec} act infinitesimally, with parameter
$\bar\epsilon_A^{\dot\alpha}$, by
\[
 \delta\mu_i^{\dot\alpha}
 =\bar\epsilon_A^{\dot\alpha}\chi_i^A,
 \qquad
 \delta\lambda_i=\delta\chi_i=0 .
\]
They preserve the fixed-period kinematics because
$\delta(\mu_{i+n}^{\dot\alpha}-\mu_i^{\dot\alpha}
-q^{\alpha\dot\alpha}\lambda_{i\alpha})
=\bar\epsilon_A^{\dot\alpha}(\chi_{i+n}^A-\chi_i^A)=0$.  The dotted part
of the amplitude $\bar Q$ generator is
\begin{equation}
 \qb^A_{\dot\alpha}
 =\sum_{i=1}^n\chi_i^A
 \frac{\partial}{\partial\mu_i^{\dot\alpha}}
 \qquad \dot\alpha=1,2 ,
 \label{eq:qbar-fixed-q}
\end{equation}
where the sum runs over one fundamental period and the derivatives act at
fixed $\lambda_i$ and $q$.  Thus, $\qb$ is a vector
field on the $\gamma=0$, fixed-$q$ periodic kinematic space.
The undotted part of $\bar Q$ does not preserve these periodicity
conditions; appendix~\ref{app:monodromy} gives its action on the
monodromy.

For the anomalous equation, we remove the common momentum and supermomentum
delta functions and separate the BDS factor.  Let $F_{n,k}$ denote the
$\mathrm N^k\mathrm{MHV}$ form factor with these delta functions stripped,
where $k=0$ is the MHV sector.  We write the MHV form factor and helicity
ratios as
\begin{equation}
 F_{n,0}=F_{n,0}^{\rm tree}D_n e^{R_n},
 \qquad
 \cR_{n,k}=\frac{F_{n,k}}{F_{n,0}},
 \qquad \cR_{n,0}=1 .
 \label{eq:normalization}
\end{equation}
Here $D_n$ is the form-factor BDS factor
\cite{Bern:2005iz,Brandhuber:2012vm}, with the dimensional-regularization
convention specified in appendix~\ref{app:monodromy}.  The logarithmic
remainder $R_n$ starts at two loops.  The ratio $\cR_{n,k}$ has Grassmann
degree $4k$.  We use
the same subtraction scheme and analytic continuation at multiplicities
$n$ and $n+1$.

\subsection{Statement of the conjecture}

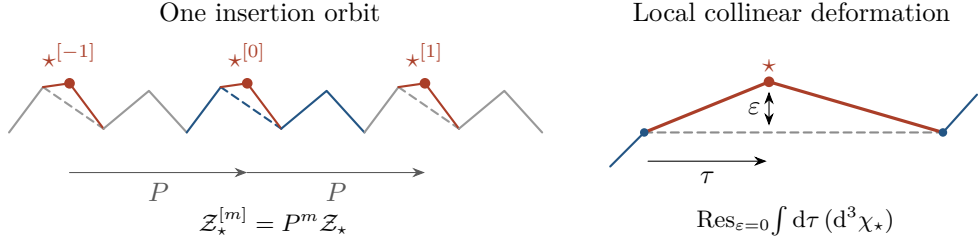
\begin{figure}[h]
\centering
\begin{tikzpicture}[x=1cm,y=1cm,>=Stealth,
  every node/.style={font=\small},line cap=round,line join=round]
  \node at (3.75,3.10) {One insertion orbit};
  \foreach \dx/\col/\lab in {.30/black!40/{-1},2.65/periodblue/{0},5.00/black!40/{1}}{
    \begin{scope}[shift={(\dx,1.50)}]
      \draw[\col,thick] (0,0)--(.45,.60);
      \draw[\col,thick,densely dashed] (.45,.60)--(1.25,.05);
      \draw[insertionred,thick] (.45,.60)--(.80,.65)--(1.25,.05);
      \draw[\col,thick] (1.25,.05)--(1.85,.55)--(2.35,0);
      \fill[insertionred] (.80,.65) circle (2pt);
      \node[above,text=insertionred] at (.80,.73) {$\star^{[\lab]}$};
    \end{scope}
  }
  \draw[->,black!65] (1.10,.97)--(3.45,.97) node[midway,below] {$P$};
  \draw[->,black!65] (3.45,.97)--(5.80,.97) node[midway,below] {$P$};
  \node[font=\footnotesize] at (3.8,.33) {$\mathcal Z_{\star}^{[m]}=P^m\mathcal Z_{\star}$};

  \node at (10.65,3.10) {Local collinear deformation};
  \coordinate (left) at (8.7,1.50);
  \coordinate (right) at (12.65,1.50);
  \coordinate (foot) at (10.35,1.50);
  \coordinate (star) at (10.35,2.17);
  \draw[black!45,thick,densely dashed] (left)--(right);
  \draw[periodblue,thick] (8.25,1.05)--(left);
  \draw[periodblue,thick] (right)--(13.10,2.0);
  \draw[insertionred,very thick] (left)--(star)--(right);
  \fill[periodblue] (left) circle (1.6pt);
  \fill[periodblue] (right) circle (1.6pt);
  \fill[insertionred] (star) circle (2pt);
  \node[above,text=insertionred] at (star) {$\star$};
  \draw[<->,thin] (10.35,1.59)--(10.35,2.04)
    node[midway,left,inner sep=3pt] {$\varepsilon$};
  \draw[->,thin] (8.75,1.12)--(10.32,1.12)
    node[midway,below] {$\tau$};
  \node[font=\footnotesize] at (10.70,.33)
    {$\operatorname*{Res}_{\varepsilon=0}\!\int\dd\tau\,(\dd^3\chi_\star)$};
\end{tikzpicture}
\caption{Collinear insertion on the periodic contour. Left: the additional
cusp and its images under the period translation $P$. Right: a schematic
local deformation, with longitudinal parameter $\tau$ and transverse
collinear parameter $\varepsilon$, as defined in \eqref{eq:collinear-curve}.}
\label{fig:qbar-insertion}
\end{figure}

Let $\cI_i$ denote the operation that inserts an additional particle collinear with edge $i$, together
with its complete periodic orbit, as illustrated in
figure~\ref{fig:qbar-insertion}.  We conjecture the all-loop MHV equation
\begin{equation}
 \boxed{
 \qb^A_{\dot\alpha}R_n
 =a(g)\sum_{i=1}^n\cI^A_{i,\dot\alpha}
 \left[
 e^{R_{n+1}-R_n}\cR_{n+1,1}
 -\cR_{n+1,1}^{(0)}
 \right],
 \qquad a(g)=\frac{\Gamma_{\rm cusp}(g)}4 .}
 \label{eq:mhv-equation}
\end{equation}
The superscript $(0)$ denotes tree level.  All quantities are evaluated
in the same subtraction scheme and with a common collinear regulator at
adjacent multiplicities.  We use
\begin{equation}
 g^2=\frac{g_{\rm YM}^2N_c}{16\pi^2},\qquad
 a(g)=g^2-2\zeta_2g^4+\mathcal O(g^6).
 \label{eq:coupling-convention}
\end{equation}
The exponential converts the NMHV ratio, normalized
by the full MHV form factor, to the BDS normalization used in the
conjecture.  The tree term subtracts the variation of the BDS reference.
Equation~\eqref{eq:mhv-equation} is understood order by order in
perturbation theory. Since $a(g)$ starts at order $g^2$, its $L$-loop
component involves form-factor data through at most $L-1$ loops.
In particular,
\begin{equation}
 \qb R_n^{(2)}=\sum_{i=1}^n\cI_i[\cR_{n+1,1}^{(1)}] .
 \label{eq:two-loop-equation}
\end{equation}
This is the relation used in the two-loop tests below.

For higher helicity sectors, define the BDS-normalized form factor
\begin{equation}
 \bR_{n,k}=\frac{F_{n,k}}{F_{n,0}^{\rm tree}D_n}
 =e^{R_n}\cR_{n,k} .
 \end{equation}
We conjecture the following extension to general helicity:
\begin{equation}
 \qb\bR_{n,k}=a(g)\sum_i\cI_i
 \left[\bR_{n+1,k+1}-\bR_{n,k}\cR_{n+1,1}^{(0)}\right].
 \label{eq:general-helicity}
\end{equation}
For $k=0$, we have $\bR_{n,0}=e^{R_n}$. Dividing
\eqref{eq:general-helicity} by $e^{R_n}$ gives
\eqref{eq:mhv-equation}, including the factor $e^{R_{n+1}-R_n}$
and the tree subtraction.
Integrating over three insertion fermions leaves Grassmann degree
$4k+1$, as required for a $\qb$ variation.  The one-loop NMHV calculation in
section~\ref{sec:one-loop-nmhv} tests this extension beyond MHV.

\subsection{Collinear insertion and finite prescription}

On the universal cover, we parametrize the collinear restriction by
\begin{equation}
 \mathcal Z_{\star i}
 =\mathcal Z_i-\varepsilon\mathcal Z_{i-1}
 +c_i\varepsilon\tau\mathcal Z_{i+1}
 +c_i'\varepsilon^2\mathcal Z_{i+2},
 \qquad
 \mathcal Z_{\star i}^{[m]}=P^m\mathcal Z_{\star i} .
 \label{eq:collinear-curve}
\end{equation}
The parameter $\varepsilon$ controls the transverse collinear limit,
while $0<\tau<\infty$ parametrizes the longitudinal direction. The
bosonic part of \eqref{eq:collinear-curve} is imposed in the integrand.
All images share one independent fermionic coordinate $\chi_\star$,
which is restricted to the curve only after the Grassmann integration
specified below. The insertion geometry is illustrated in
figure~\ref{fig:qbar-insertion}.

To leading order, the insertion is
\begin{equation}
 \cI^A_{i,a}[X]
 =\Res_{\varepsilon=0}\int_0^\infty
 (\dd^{2|3}\mathcal Z_{\star i})^A_a X,
 \qquad
 (\dd^{2|3}\mathcal Z_{\star i})^A_a
 =c_i(i-1,i,i+1)_a\,
 \varepsilon\dd\varepsilon\dd\tau\,(\dd^3\chi_\star)^A .
 \label{eq:insertion}
\end{equation}
Here $a$ is a bosonic twistor index; the dotted components in
\eqref{eq:mhv-equation} correspond to the $\mu$ sector.  We use
\begin{equation}
 (ijk)_a=\epsilon_{abcd}Z_i^bZ_j^cZ_k^d,\qquad
 (\dd^3\chi)^A=\frac{\epsilon^{ABCD}}{3!}
 \dd\chi^B\dd\chi^C\dd\chi^D .
 \label{eq:insertion-measure}
\end{equation}
The factors $c_i,c_i'$ specify the collinear parametrization discussed in
appendix~\ref{app:monodromy}.
The Grassmann integrations are performed before restricting the remaining
fermionic coordinate to the collinear curve. The residue in
\eqref{eq:insertion} is understood on the complete regulated longitudinal
integral. In practice, we extract the terms proportional to
$\dd\varepsilon/\varepsilon$ in the collinear expansion, retaining any
powers of $\log\varepsilon$ until after the longitudinal integration.
A finite residue requires these logarithmic terms to cancel in the
integrated result, as verified in the perturbative examples below.

The two terms in \eqref{eq:mhv-equation} must be integrated together,
using the same longitudinal coordinate and endpoint cutoff.  Their separate
$\dd\tau/\tau$ singularities cancel by the common soft and collinear limits.
The periodic image sum is performed before taking the residue.  Thus, near
the boundary of a chosen period, the neighbouring cusp in the local
factorization is included through its translated image.

The right-hand side has the interpretation of a quantum correction to the
$\qb$ action, as in the amplitude anomaly equation.  Summing one
insertion orbit over the $n$ edges preserves cyclicity.  In the MHV equation
the tree terms in the bracket cancel, while $a(g)$ begins at one loop;
the first nonzero relation is therefore \eqref{eq:two-loop-equation}, in
accord with the normalization of $R_n$.

The two-point case gives a simple example of the fermionic projection.
At vanishing fermionic monodromy, a fermionic translation sets both external
$\chi$'s and all their images to zero.  The degree-four three-point source
is then proportional to $\prod_{A=1}^4\chi_\star^A$.  Three Grassmann
derivatives leave one factor $\chi_\star^A$, which vanishes on the collinear
curve.  Each edge thus gives zero before the $\varepsilon$ residue or
$\tau$ integration.  The equation reduces to $0=0$, so the two-point case
places no constraint on the finite normalization of the insertion.

\section{Completing the MHV differential with parity}
\label{sec:differential}

To reconstruct the MHV remainder from \eqref{eq:mhv-equation}, we need
its derivatives with respect to independent kinematic variables. The
$\qb$ equation initially gives only derivatives that vary the dotted
spinors while keeping the undotted spinors and the operator momentum
fixed. At three points these determine the whole differential. At four
points they determine four derivatives, whereas the remainder depends
on five variables. Parity supplies the missing derivative by relating a
variation of the undotted spinors to a dotted variation at a different
kinematic point.

We first extract the derivatives from the collinear integral, then
combine them with its parity-transformed evaluation to obtain the full
differential. A four-point example makes this construction explicit.
Throughout this section, the kinematics are complexified, $n\geq3$,
and $q$ is fixed with $q^2\neq0$. We work at generic points and require
the spinor brackets appearing in denominators to be nonzero.

\subsection{Derivatives supplied by the collinear integral}

The generator in \eqref{eq:qbar-fixed-q} differentiates with respect to
$\mu_i$, so the first step is to express its action in spinor-helicity
variables. Inverting the incidence relation of section~\ref{sec:equation}
gives
\begin{equation}
 \widetilde\lambda_i=
 \frac{\mu_{i-1}}{\br{i-1,i}}
 -\frac{\br{i-1,i+1}\mu_i}{\br{i-1,i}\br{i,i+1}}
 +\frac{\mu_{i+1}}{\br{i,i+1}} .
 \label{eq:inverse-incidence-differential}
\end{equation}
Indices outside the chosen period denote periodic images, including
their $q$-dependent shifts. The same linear combination of the
fermionic coordinates $\chi_j^A$ gives the on-shell Grassmann variables
$\eta_i^A$; here the images obey $\chi_{j+n}^A=\chi_j^A$, with no
fermionic shift on the $\gamma=0$ slice. The chain rule then gives\footnote{In on-shell
variables, this is the ordinary superconformal generator
$\bar s_{\dot\alpha}^{A}$~\cite{Bullimore:2011kg,Caron-Huot:2011dec}.}
\begin{equation}
 \qb^A_{\dot\alpha}R_n
 =\sum_i\eta_i^A
 \frac{\partial R_n}{\partial\widetilde\lambda_i^{\dot\alpha}},
 \qquad
 \sum_i\lambda_i^\alpha\eta_i^A=0 .
 \label{eq:incidence-fermions}
\end{equation}
The second relation is the zero-supermomentum condition. It ensures
that the combined variation on the right preserves
$q=\sum_i\lambda_i\widetilde\lambda_i^T$.

The Grassmann variables in \eqref{eq:incidence-fermions} are therefore not all
independent. Choose two reference legs $a,b$ with $\br{ab}\neq0$ and
eliminate $\eta_a^A,\eta_b^A$ using the second relation. Express the
collinear integral on the right-hand side of \eqref{eq:mhv-equation}
in the remaining Grassmann variables, and denote their coefficients by
$F_{r\dot\alpha}$. The equation becomes
\begin{equation}
 \qb^A_{\dot\alpha}R_n
 =\sum_{r\neq a,b}\eta_r^A F_{r\dot\alpha},
 \qquad
 F_{r\dot\alpha}=\mathsf A_{r\dot\alpha}R_n,
 \label{eq:source-coefficients}
\end{equation}
where
\begin{equation}
 \mathsf A_{r\dot\alpha}
 =\frac{\partial}{\partial\widetilde\lambda_r^{\dot\alpha}}
 -\frac{\br{rb}}{\br{ab}}
 \frac{\partial}{\partial\widetilde\lambda_a^{\dot\alpha}}
 -\frac{\br{ar}}{\br{ab}}
 \frac{\partial}{\partial\widetilde\lambda_b^{\dot\alpha}},
 \qquad r\neq a,b .
 \label{eq:A-derivatives}
\end{equation}
The first term in $\mathsf A_{r\dot\alpha}$ varies the dotted spinor
of leg $r$; the other two adjust those of the reference legs $a,b$
so that $\sum_i\lambda_i\delta\widetilde\lambda_i^T=0$.
Thus $F_{r\dot\alpha}$ gives a derivative of $R_n$ at fixed $\lambda$
and $q$. There are $2n-4$ such derivatives.

Rewriting the collinear integral in terms of $\eta$ requires one
consistency property. A common shift of the origin of the dual Grassmann coordinates,
$\chi_i^A\mapsto\chi_i^A+\lambda_{i\alpha}\kappa^{\alpha A}$, leaves
all $\eta_i^A$ unchanged; here $\kappa^{\alpha A}$ is a constant
Grassmann parameter. At generic kinematics these are the only shifts
that leave all $\eta_i^A$ unchanged. The left-hand side of \eqref{eq:mhv-equation} is invariant
under this shift because $R_n$ is invariant under translations of the
dual coordinates. The collinear integral must have the same invariance.
Once it does, its coefficients in the independent $\eta_r^A$ give the
$F_{r\dot\alpha}$ without retaining the two redundant combinations
of $\chi_i^A$.

\subsection{Parity and the full differential}

Parity exchanges dotted and undotted spinors. Since the operator
momentum is held fixed, this exchange must be followed by a Lorentz
transformation that restores $q$. To write the transformation explicitly,
collect the spinors into a $2\times n$ matrix $\Lambda$ and an
$n\times2$ matrix $V$:
\begin{equation}
 \Lambda=(\lambda_1\ \cdots\ \lambda_n),\qquad
 V=\begin{pmatrix}
 \widetilde\lambda_1^T\\[-1mm]\vdots\\
 \widetilde\lambda_n^T
 \end{pmatrix},\qquad \Lambda V=q .
 \label{eq:matrix-kinematics}
\end{equation}
Choose a branch of $m=\sqrt{\det q}$. The parity transformation at fixed $q$ is
\begin{equation}
 \Phi_q:(\Lambda,V)\longmapsto
 \left(\frac{q}{m}V^T,\;m\Lambda^Tq^{-T}\right),
 \qquad \Phi_q^2=1 ,
 \label{eq:fixed-q-parity}
\end{equation}
where $q^{-T}=(q^{-1})^T$. The product of the transformed spinor
matrices is again $q$.

The parity property used here is that of the scalar MHV remainder.
In the Wilson-loop description, $R_n$ is the finite part of the
logarithm of the bosonic periodic Wilson-loop expectation value after
BDS subtraction. Parity maps the contour to its parity image without
changing this expectation value, and the compensating Lorentz
transformation also leaves it invariant. With a parity-preserving
regulator and subtraction scheme, and on matched analytic branches,
\begin{equation}
 R_n\!\left(\Phi_q(\Lambda,V)\right)=R_n(\Lambda,V) .
 \label{eq:parity-invariance}
\end{equation}
The chiral NMHV form factor in the collinear integral need not itself
be parity invariant.

We can now obtain undotted derivatives by evaluating the same
collinear integral at $\Phi_q(\Lambda,V)$. For the same reference legs, with
$[ab]\neq0$, define the derivatives that vary $\lambda_r$ while
compensating with $\lambda_a,\lambda_b$:
\begin{equation}
 \mathsf B_{r\alpha}
 =\frac{\partial}{\partial\lambda_r^\alpha}
 -\frac{[rb]}{[ab]}\frac{\partial}{\partial\lambda_a^\alpha}
 -\frac{[ar]}{[ab]}\frac{\partial}{\partial\lambda_b^\alpha},
 \qquad r\neq a,b .
 \label{eq:B-derivatives}
\end{equation}
Under \eqref{eq:fixed-q-parity}, a variation of $\lambda_r$ becomes
$\delta\widetilde\lambda'_r=mq^{-1}\delta\lambda_r$, and the
angle-bracket ratios at $\Phi_q(\Lambda,V)$ become the square-bracket ratios at
the original kinematic point. Differentiating \eqref{eq:parity-invariance} therefore gives
\begin{equation}
 \mathsf B_{r\alpha}R_n=G_{r\alpha},\qquad
 G_{r\alpha}:=
 \sum_{\dot\beta}(mq^{-T})_{\alpha\dot\beta}
 F_{r\dot\beta}\!\left(\Phi_q(\Lambda,V)\right) .
 \label{eq:parity-pullback}
\end{equation}
Thus $F$ supplies the dotted derivatives and $G$ the undotted ones.
The matrix multiplying $F\!\left(\Phi_q(\Lambda,V)\right)$ is the chain-rule factor needed
to express the parity-derived derivatives at the original kinematic point.
The additional information therefore comes from a second evaluation of
the source, at the parity-related point.

To combine these results, consider any infinitesimal change of the
spinors,
\begin{equation}
 (\delta\Lambda)V+\Lambda(\delta V)=0 .
 \label{eq:fixed-q-variation}
\end{equation}
This is simply the condition $\delta q=0$. We want the variation of
$R_n$ along this change. The coefficients $F$ determine it when $\delta\Lambda=0$,
and $G$ determine it when $\delta V=0$. For a general change, set
\begin{equation}
 \begin{aligned}
 \delta\Lambda_0&=\delta\Lambda-(\delta\Lambda)Vq^{-1}\Lambda,\\
 \delta V_0&=\delta V+Vq^{-1}(\delta\Lambda)V .
 \end{aligned}
 \label{eq:tangent-split}
\end{equation}
These adjusted variations satisfy $(\delta\Lambda_0)V=0$ and $\Lambda(\delta V_0)=0$.
They differ from the original variation only by little-group and
Lorentz transformations, which do not change the scalar remainder,
as shown below. Provided the two evaluations agree on the variations
they have in common, the collinear integral equation predicts the full variation
\begin{equation}
 \boxed{
 \delta R_n=
 \sum_{r\neq a,b}\left[
 \sum_{\alpha}(\delta\Lambda_0)_{\alpha r}G_{r\alpha}
 +\sum_{\dot\alpha}(\delta V_0)_{r\dot\alpha}F_{r\dot\alpha}
 \right].}
 \label{eq:complete-differential}
\end{equation}
Formula~\eqref{eq:complete-differential}
can be used in any coordinate system: for each coordinate, take $\delta\Lambda,\delta V$
to be the derivatives of the spinor matrices with respect to that
coordinate. The formula then gives the corresponding derivative of
$R_n$ directly.

To justify the adjustment in \eqref{eq:tangent-split}, recall that
little-group rescalings and Lorentz transformations preserving $q$
act infinitesimally as
\begin{equation}
 (\Lambda T,-TV),\qquad
 (H\Lambda,-Vq^{-1}Hq),\qquad
 T\ \hbox{diagonal},\quad \operatorname{tr}H=0 .
 \label{eq:kinematic-gauge-directions}
\end{equation}
The variation removed in \eqref{eq:tangent-split} is of this form:
the traceless part of $(\delta\Lambda)Vq^{-1}$ generates a Lorentz
transformation, while its scalar part generates a common little-group rescaling. It therefore
annihilates $R_n$. The contractions with $F$ and $G$ in
\eqref{eq:complete-differential} must likewise be invariant under
changes of spinor representatives; this is the covariance condition
on the coefficients computed from the collinear integral.

The same argument explains why parity supplies exactly the missing
physical derivatives. Momentum conservation leaves $4n-4$ spinor
components. Removing the $n$ little-group rescalings and the three
Lorentz transformations preserving $q$ leaves $3n-7$ physical variables.
Neither set of chiral variations contains a nonzero transformation
of \eqref{eq:kinematic-gauge-directions} at a generic point. Indeed, a
transformation that leaves all $\lambda_i$ fixed would make them all
eigenvectors of the same traceless $2\times2$ matrix. Three distinct
projective directions force that matrix, and hence the transformation,
to vanish. The dotted argument is identical. Each set therefore
provides $2n-4$ independent directions. Since
\eqref{eq:tangent-split} expresses every physical variation in terms
of the two sets, parity adds the remaining $n-3$ directions: none at
three points, one at four points, and two at five points.
At four points, the two sets of four derivatives share three physical
directions, so only one direction supplied by $G$ is new.

There are also $2(2n-4)-(3n-7)=n-1$ directions that both evaluations
describe. They must give the same derivative in these directions.
An explicit way to impose this condition is to take a diagonal matrix
$T$ and form
\begin{equation}
 \delta_T\Lambda=\Lambda T(1_n-Vq^{-1}\Lambda),\qquad
 \delta_T V=(1_n-Vq^{-1}\Lambda)TV .
 \label{eq:chiral-overlap}
\end{equation}
Here $(\delta_T\Lambda)V=0$ and $\Lambda(\delta_T V)=0$. The difference between the
undotted variation $(\delta_T\Lambda,0)$ and the dotted variation $(0,\delta_T V)$ is
\begin{equation}
 (\delta_T\Lambda,-\delta_T V)=(\Lambda T,-TV)
 -(\Lambda TVq^{-1}\Lambda,-Vq^{-1}\Lambda TV) .
 \label{eq:overlap-gauge-identity}
\end{equation}
Both terms on the right leave the remainder invariant by the preceding
argument, so the two variations describe the same physical change.
A scalar $T$ gives zero. Conversely, $\delta_T\Lambda=0$ forces $T$ to be scalar
at a generic point, by the same eigenvector argument. Diagonal $T$
modulo its scalar part therefore parametrizes all $n-1$ common
directions. The required agreement is
\begin{equation}
 \sum_{r\neq a,b}\sum_\alpha(\delta_T\Lambda)_{\alpha r}G_{r\alpha}
 =\sum_{r\neq a,b}\sum_{\dot\alpha}(\delta_T V)_{r\dot\alpha}F_{r\dot\alpha}
 \qquad\hbox{for every diagonal }T .
 \label{eq:overlap-compatibility}
\end{equation}
This condition makes \eqref{eq:complete-differential} independent
of how a physical variation is split between the two chiralities.
When $F,G$ are computed from the conjectured collinear integral, the
resulting coordinate derivatives must also satisfy equality of mixed
partial derivatives. On a simply connected patch, this ensures that
they can be integrated to $R_n$ up to an additive constant fixed by boundary data.

\subsection{A four-point example}
\label{sec:four-point-differential}

At four points, the role of parity is to determine a single derivative
that the original evaluation leaves undetermined. Choose a spinor
frame adapted to $q$ and measure momenta in units of $m$, so that
$q=1_2$ and $m=+1$. A useful local choice of spinors is
\begin{equation}
 \Lambda=\begin{pmatrix}1&0&1&1\\0&1&1&\zeta\end{pmatrix},\qquad
 V=\begin{pmatrix}
 1-a-c&-b-d\\
 -a-\zeta c&1-b-\zeta d\\
 a&b\\c&d
 \end{pmatrix} .
 \label{eq:four-point-differential-chart}
\end{equation}
Momentum conservation $\Lambda V=1_2$ is automatic. The five
independent coordinates are $(a,b,c,d,\zeta)$; here $a,b$ are coordinate
names. We use legs 1 and 2 as the reference pair and require $[12]\neq0$.
The first three undotted spinors have fixed projective directions,
so $\zeta$ describes their cross-ratio with the fourth. Changing any
of $a,b,c,d$ keeps $\Lambda$ fixed and adjusts the first two rows of
$V$ to preserve $q$. The original collinear integral therefore gives,
writing $R=R_4$,
\begin{equation}
 \partial_aR=F_{3\dot1},\qquad
 \partial_bR=F_{3\dot2},\qquad
 \partial_cR=F_{4\dot1},\qquad
 \partial_dR=F_{4\dot2} .
 \label{eq:four-point-chiral-coordinates}
\end{equation}

The missing derivative is $\partial_\zeta R$. Varying $\zeta$ at fixed
$a,b,c,d$ changes both $\Lambda$ and $V$, so this derivative involves
both the undotted data $G$ and the dotted data $F$. For this variation,
take $\delta\Lambda=\partial_\zeta\Lambda$ and
$\delta V=\partial_\zeta V$. The only nonzero entry of
$\delta\Lambda$ is $(\delta\Lambda)_{24}=1$, while only the second
row of $\delta V$ is nonzero, equal to $(-c,-d)$.
Using \eqref{eq:tangent-split}, the components needed in
\eqref{eq:complete-differential} are
\begin{equation}
 (\delta\Lambda_0)_{\bullet3}=\binom{0}{-c-d},\qquad
 (\delta\Lambda_0)_{\bullet4}=\binom{0}{1-c-\zeta d},\qquad
 (\delta V_0)_{\{3,4\}\bullet}
 =\begin{pmatrix}bc&bd\\cd&d^2\end{pmatrix} .
 \label{eq:four-point-split-components}
\end{equation}
Here a bullet denotes the whole column or row; for example,
$(\delta\Lambda_0)_{\bullet3}$ is the third column of $\delta\Lambda_0$. Substitution gives
\begin{equation}
 \begin{split}
 \partial_\zeta R={}&-(c+d)G_{32}
 +(1-c-\zeta d)G_{42}\\
 &+bcF_{3\dot1}+bdF_{3\dot2}
 +cdF_{4\dot1}+d^2F_{4\dot2} .
 \end{split}
 \label{eq:four-point-fifth-derivative}
\end{equation}
Since $mq^{-T}=1_2$ in this frame, $G_{r2}=F_{r\dot2}\!\left(\Phi_q(\Lambda,V)\right)$.
Thus the first two terms are obtained by evaluating the same collinear
integral at $\Phi_q(\Lambda,V)=(V^T,\Lambda^T)$, with reference legs 1 and 2;
the remaining four terms use its original evaluation. The coefficients
at the parity point are defined in this displayed frame. If that frame
is changed, the coefficients must be transformed accordingly.

Together, \eqref{eq:four-point-chiral-coordinates} and
\eqref{eq:four-point-fifth-derivative} determine all five derivatives
of $R_4$. Derivatives with respect to independent Mandelstam invariants
then follow by the chain rule.

\section{Perturbative evidence}
\label{sec:evidence}

We apply the periodic equation to one-loop NMHV and two-loop MHV form
factors.  The NMHV calculation tests the helicity extension with a
nontrivial tree-level source.  In the MHV sector, the three-point example
shows how the regulated one-loop source determines a complete two-loop
function once its collinear boundary value is specified.  Four- and five-point
comparisons extend these applications to larger kinematic spaces.

\subsection{One-loop NMHV form factors}
\label{sec:one-loop-nmhv}

The first possible one-loop NMHV equation has three points on the left and
four-point tree data on the right.  At $\gamma=0$, the one-loop correction
to the three-point NMHV ratio vanishes.  On each edge, both
$\cR_{4,2}^{(0)}$ and the subtraction
$\cR_{3,1}^{(0)}\cR_{4,1}^{(0)}$ also vanish after the Grassmann projection
and fermionic pullback, before taking the collinear residue.  This
provides a degenerate consistency check; the first nontrivial test of
the bosonic functions occurs at four points.

The four-point equation is
\begin{equation}
 \qb\cR_{4,1}^{(1)}
 =\sum_{i=1}^4\cI_i
 \left[\cR_{5,2}^{(0)}
 -\cR_{4,1}^{(0)}\cR_{5,1}^{(0)}\right].
 \label{eq:one-loop-nmhv}
\end{equation}
We construct its two sides independently.  The left-hand input is the
integrated one-loop NMHV form factor obtained from generalized
unitarity~\cite{Bork:2012tt,Bianchi:2018rrj}, with the MHV normalization
of \eqref{eq:normalization}.  Define the cyclic invariants
\begin{equation}
 u_i=\frac{s_{i,i+1}}{q^2},\qquad
 v_i=\frac{s_{i,i+1,i+2}}{q^2},\qquad i\pmod4 .
 \label{eq:four-point-invariants}
\end{equation}
In terms of reduced finite scalar integrals, the complete input is
\begin{align}
 \cR_{4,1}^{(1)}={}&\sum_{i=1}^4 c_i^{1\mathrm m}
 \Bigl[F^{1\mathrm m}(u_i,u_{i+1};v_i)
 +F^{2\mathrm{mh}}(u_{i+2},v_i;u_i,1)
 +F^{2\mathrm{mh}}(u_{i+3},v_i;u_{i+1},1)\Bigr]\nonumber\\
 &-2\Bigl[c_1^{3\mathrm m}F^{3\mathrm m}(u_3,u_1)
          +c_2^{3\mathrm m}F^{3\mathrm m}(u_4,u_2)\Bigr]
 -\cR_{4,1}^{(0)}\mathcal M_{4,\mathrm{fin}}^{(1)}.
 \label{eq:four-point-nmhv-input}
\end{align}
Here $c_i^{1\mathrm m}$ is the Grassmann coefficient shared by the three
displayed box functions, and $\mathcal M_{4,\mathrm{fin}}^{(1)}$ is the
finite part of the one-loop coefficient $\mathcal M_4^{(1)}(\epsilon)$
in the MHV ratio $\mathcal M_n=F_{n,0}/F_{n,0}^{\rm tree}$ defined in
appendix~\ref{app:monodromy}, obtained by removing the unexpanded
infrared power terms.  The function $F^{3\mathrm m}$ is the three-mass
triangle function, with the indicated cross ratios as its
arguments~\cite{Bianchi:2018rrj}.
The coefficients $c_1^{3\mathrm m}$ and $c_2^{3\mathrm m}$ label the two
triangle channels in the order shown and include both solutions of
the triple cut.
The explicit inputs and details of the component comparison are collected
in appendix~\ref{app:one-loop-nmhv-comparison}.
For this representation, the complete Grassmann coefficients satisfy
$\qb c_i^{1\mathrm m}=\qb c_1^{3\mathrm m}
=\qb c_2^{3\mathrm m}=\qb\cR_{4,1}^{(0)}=0$.  Thus $\qb$ acts
only on the scalar functions in \eqref{eq:four-point-nmhv-input}.
Differentiating the boxes and triangles reduces the result to rational
coefficients multiplying $\log u_i$ and $\log v_i$.

On the right we use the periodic BCFW representations of
$\cR_{5,2}^{(0)}$, $\cR_{5,1}^{(0)}$ and $\cR_{4,1}^{(0)}$
\cite{Bork:2014eqa}.  For each insertion edge
$e=1,\ldots,4$, the five-point functions are evaluated on the
corresponding periodic insertion curve, while the four-point factor
retains the original external configuration.  We first integrate the
three insertion fermions, then restrict the remaining fermion to the
curve and take the collinear residue.  After stripping the dotted
measure factor in \eqref{eq:insertion}, the four Grassmann structures
on each edge give rational kernels
\begin{equation}
 k_e^{(r)}(\tau)=\sum_{a\in\mathcal A_e}
 \frac{b_{e,a}^{(r)}}{\tau+a},\qquad
 \sum_a b_{e,a}^{(r)}=0,\qquad r=0,1,2,3 .
 \label{eq:simple-pole-kernel}
\end{equation}
Here $r$ labels the four Grassmann structures.  The only possible poles are
at $\tau=-a$, with
\begin{equation}
 \mathcal A_e=
 \left\{\frac{u_{e+2}}{v_{e+2}},\,
 \frac{v_e}{u_{e+1}},\,v_{e+1},\,\frac1{v_{e+1}}\right\}.
 \label{eq:four-point-poles}
\end{equation}
The residue sum in \eqref{eq:simple-pole-kernel} cancels the logarithm
at the upper endpoint.  With the common endpoint prescription, the
longitudinal integral is therefore obtained analytically as
\begin{equation}
 \int_0^\infty k_e^{(r)}(\tau)\dd\tau
 =-\sum_{a\in\mathcal A_e}b_{e,a}^{(r)}\log a .
 \label{eq:logarithmic-integral}
\end{equation}
Restoring the dotted measure and summing the four edges reproduces
$\qb\cR_{4,1}^{(1)}$ in all eight independent dotted Grassmann components.
Both calculations keep five independent bosonic coordinates symbolic and
give identical rational coefficients of $\log u_i$ and $\log v_i$, with
no additional rational term.

\subsection{Two-loop MHV form factors}
\label{sec:three-point-reconstruction}
\label{sec:higher-point-tests}

At three points, define
\begin{equation}
 u=\frac{s_{12}}{q^2},\qquad
 v=\frac{s_{23}}{q^2},\qquad
 w=1-u-v,
 \label{eq:uvw}
\end{equation}
and initially work in the positive chamber $u,v,w>0$.  A fermionic origin
can be chosen such that $\chi_1=\chi_2=0$.  In the three-point twistor
patch of \eqref{eq:mellin-twistor-patch},
\begin{equation}
 \qb^A_{\dot1}R_3=\chi_3^A\partial_uR_3,
 \qquad
 \qb^A_{\dot2}R_3=\chi_3^A(-\partial_u+\partial_v)R_3 .
 \label{eq:three-point-qbar}
\end{equation}
Equation~\eqref{eq:three-point-qbar} allows us to extract the two
independent derivatives of $R_3$ from the coefficients of $\chi_3^A$
in the collinear source.

We first evaluate the third insertion edge,
$\cI_3[\cR_{4,1}^{(1)}]$, using the four-point one-loop NMHV ratio in
\eqref{eq:four-point-nmhv-input}.  After the Grassmann integration,
collinear expansion and longitudinal integration, its two dotted components
are $\chi_3^A vK(u,v,w)$ and $\chi_3^A wK(u,v,w)$, respectively.
The evaluation of $K$ and the subsequent reconstruction of the remainder
are detailed in
appendix~\ref{app:three-point-kernel}.
Using \eqref{eq:three-point-qbar}, this edge contributes
\begin{equation}
 \omega_{\qb,3}
 =K(u,v,w)\,[v\,\dd u+(v+w)\,\dd v]
 =vwK(u,v,w)\,\dlog\frac{v}{w} .
 \label{eq:one-edge-three-point-form}
\end{equation}
Adding the cyclic contributions from the other two edges yields
\begin{align}
 D_u={}&vK(u,v,w)-(u+w)K(v,w,u)+vK(w,u,v),\\
 D_v={}&(v+w)K(u,v,w)-uK(v,w,u)-uK(w,u,v),
 \label{eq:three-point-D}
\end{align}
and hence
\begin{equation}
 \omega_{\qb}=D_u\,\dd u+D_v\,\dd v .
 \label{eq:three-point-one-form}
\end{equation}

To integrate this one-form, introduce the positive coordinates
\begin{equation}
 u=\frac{x}{1+x+y},\qquad
 v=\frac{y}{1+x+y},\qquad
 w=\frac{1}{1+x+y} .
 \label{eq:positive-chart}
\end{equation}
After rationalizing the triangle roots, its pullback belongs to a
hyperlogarithmic function space with $x$ letters
$\{0,-1,-y,-1-y\}$ and $y$ letters $\{0,-1\}$.  The limit $x\to0$
corresponds to the $u\to0$ collinear boundary. The BDS collinear condition
$R_3^{(2)}\to R_2^{(2)}=0$, with the chosen finite Sudakov normalization,
fixes the additive constant. Integrating the $x$ component with this
vanishing boundary value gives a weight-four
function whose $y$ derivative also reproduces the source.  The result
agrees with the known two-loop remainder, including all beyond-symbol
terms~\cite{Brandhuber:2012vm,Dixon:2020bbt}.  Writing
$(\nu_1,\nu_2,\nu_3)=(u,v,w)$ and $\nu_{i+3}=\nu_i$, it is
\begin{align}
 R_3^{(2)}={}&-2\sum_{i=1}^3\left[
 J_4\!\left(-\frac{\nu_i\nu_{i+1}}{\nu_{i+2}}\right)
 +4\Li_4(1-1/\nu_i)+\frac{\log^4\nu_i}{6}\right]
 -\frac{\log^4(\nu_1\nu_2\nu_3)}{24}\nonumber\\
 &-2\left[\sum_{i=1}^3\Li_2(1-1/\nu_i)\right]^2
 +\frac12\left[\sum_{i=1}^3\log^2\nu_i\right]^2
 -\frac{23}{2}\zeta_4,
 \label{eq:explicit-three-point-remainder}
\end{align}
where
\begin{equation}
 J_4(z)=\Li_4(z)-\log(-z)\Li_3(z)
 +\frac12\log^2(-z)\Li_2(z)
 -\frac16\log^3(-z)\Li_1(z)-\frac1{48}\log^4(-z).
 \label{eq:J4-definition}
\end{equation}
Here $\Li_1(z)=-\log(1-z)$; all branches are fixed by continuation from
$u,v,w>0$.

At higher multiplicity we compare the chiral differential directly.  For
four points, the source in \eqref{eq:two-loop-equation} is the five-point
one-loop NMHV ratio, with box coefficients from generalized unitarity and
the three-mass triangle leading singularities of ref.~\cite{Bianchi:2018rrj}.
In the complete integrand, spurious poles on the integration contour cancel,
and collinear regulator logarithms give no contribution after integration
with the common endpoint cutoff. We choose the nondegenerate rational point
\begin{equation}
 \begin{aligned}
 Z_1&=(1,0,0,0),&\qquad Z_2&=(0,1,0,0),\\
 Z_3&=(1,1,\tfrac12,\tfrac3{10}),& Z_4&=(1,2,\tfrac14,\tfrac45),
 \end{aligned}
 \qquad Q=\mathbf1_2.
 \label{eq:evidence-rational-point}
\end{equation}
Evaluating the collinear integral at this point gives the four derivatives
of $R_4^{(2)}$ with respect to
$\mu_3^{\dot1},\mu_3^{\dot2},\mu_4^{\dot1},\mu_4^{\dot2}$.
As shown in table~\ref{tab:evidence-four-point-mu}, these agree with the
derivatives obtained from the function-level result of
ref.~\cite{Dixon:2024yvq}. This comparison includes terms beyond the symbol.
Varying the integration range and quadrature order gives changes of
approximately $1.6\times10^{-19}$ under the final refinement.

\begin{table}[htbp]
\centering
\small
\begin{tabular}{crr}
\toprule
derivative & independent target & source minus target \\
\midrule
$\partial_{\mu_3^{\dot1}}R_4^{(2)}$ & $-0.721725848262208$ & $8.82\times10^{-25}$ \\
$\partial_{\mu_3^{\dot2}}R_4^{(2)}$ & $32.581468808020706$ & $2.47\times10^{-24}$ \\
$\partial_{\mu_4^{\dot1}}R_4^{(2)}$ & $-6.962187718880099$ & $-8.90\times10^{-24}$ \\
$\partial_{\mu_4^{\dot2}}R_4^{(2)}$ & $-11.603390197634700$ & $-2.46\times10^{-24}$ \\
\bottomrule
\end{tabular}
\caption{Four-point comparison at the kinematic point
\eqref{eq:evidence-rational-point}. The third column gives the difference
between the collinear integral and the independent function-level
derivative, evaluated before rounding.}
\label{tab:evidence-four-point-mu}
\end{table}

For five points, the source is assembled from the complete six-point
one-loop NMHV ratio, including its box and triangle contributions.  We
compare with the two-loop MHV remainder symbol of ref.~\cite{Li:2024rkq}
on the one-parameter family with $Z_1,\ldots,Z_4$ and $P$ as in
\eqref{eq:evidence-rational-point} and
\begin{equation}
 Z_5(X)=\left(\frac{43}{45},\frac{1771}{900},\frac29,
                  \frac{178}{225}X\right) .
 \label{eq:evidence-five-point-family}
\end{equation}
Only $\mu_5^{\dot2}$ varies with $X$. The reference configuration at $X=1$
is nondegenerate, and the analytic branches are fixed in its neighborhood.
The derivatives are taken before restricting to this family, so
all six independent fixed-$q$ chiral components are retained, in the order
$(\mu_3^{\dot1},\mu_3^{\dot2},\mu_4^{\dot1},\mu_4^{\dot2},
\mu_5^{\dot1},\mu_5^{\dot2})$. After
integrating the five-edge source with a common endpoint cutoff, all six
components agree analytically after projecting out symbol letters that
are constant along this family.

We also test the parity reconstruction of
section~\ref{sec:four-point-differential} at the four-point configuration
\eqref{eq:evidence-rational-point}. After converting the parity result to
derivatives at the original point using \eqref{eq:parity-pullback}, all four
parity components agree with the
independent function-level differential within the observed integration
variation. They reproduce the three common directions and supply the
additional direction needed to determine the full differential at this
point. At five points, we evaluate the collinear integrals at $X=1$ in
\eqref{eq:evidence-five-point-family} and at its parity image. After the
same conversion, the two sets of six components agree on their four common
directions within the observed integration variation. The two sets of
derivatives therefore give a consistent numerical reconstruction of the
full differential at this point.

\subsection{Implications for MHV last entries}
\label{sec:mhv-last-entries}

The NMHV sources used above suggest restrictions on the last entries of
MHV remainder symbols. The amplitude $\bar Q$ equation identifies
last-entry candidates through factors of the form $\bar Q\log L$
\cite{Caron-Huot:2011dec,He:2020vob}. We apply the same strategy to
the periodic $\qb$ equation.
We assume that, at arbitrary multiplicity and loop
order, the NMHV source can be expanded in three classes of prefactor:
standard five-brackets, five-brackets multiplied by rational functions of
four-brackets, and three-mass triangle leading singularities.
For representatives of the first two classes, the
Grassmann integral and collinear residue give a longitudinal
$\dd_\tau\log$ kernel multiplying $\qb\log L(Z)$, where $L$ is rational
in the original momentum twistors and independent of $\tau$.
For the triangle class, rationalizing the cut parameter and combining
the two cut solutions gives a sum of integration kernels multiplying
$\qb\log L(Z)$, with $L(Z)$ rational and independent of the integration
variable. These $\qb\log L(Z)$ factors can therefore be taken outside
the longitudinal integral,
providing rational last-entry candidates. The calculations are detailed
in appendix~\ref{app:mhv-last-entries}.

We consider candidates of the form
\begin{equation}
 \qb\log\frac{\mathcal L_{ij}}{\mathcal L_{ik}},\qquad
 \mathcal L_{ij}=\br{i-1,i,i+1,j},\qquad
 \mathcal L_{i+n,j+n}=\mathcal L_{ij}.
 \label{eq:last-entry-periodic-brackets}
\end{equation}
All labels lie on the periodic cover. Both $j$ and $k$ obey
$2\leq|j-i|,|k-i|\leq n$, with one representative $i$ per period; this bound
on image separation is an additional assumption. There are then
$n(2n-2)$ brackets. We count logarithmic forms that are linearly
independent over constant rational coefficients. Products of the ratios in
\eqref{eq:last-entry-periodic-brackets} have zero total exponent for each
fixed first index. Together with invariance under rescaling each external
twistor, this imposes $2n-1$ independent constraints.
The brackets also satisfy
\begin{equation}
 \mathcal L_{i,i+n}=-\mathcal L_{i,i-n},\qquad
 \mathcal L_{i,i+2}=-\mathcal L_{i+1,i-1} .
 \label{eq:last-entry-bracket-identities}
\end{equation}
Within the constrained space, the first identity removes $n$ directions,
while the cyclic product of the second removes one further direction.
If these exhaust the multiplicative relations, the candidate space
therefore has dimension
\begin{equation}
 N_n^{\rm cand}=n(2n-2)-(2n-1)-(n+1)=n(2n-5),\qquad n\geq3.
 \label{eq:last-entry-count}
\end{equation}
For $n=3,\ldots,8$, explicit checks at generic kinematics give
candidate-space dimensions $3,12,25,42,63,88$, in agreement with
\eqref{eq:last-entry-count}.

At three points, the candidates can be represented by
\begin{equation}
 \dlog\frac{u}{1-u},\qquad
 \dlog\frac{v}{1-v},\qquad
 \dlog\frac{w}{1-w},\qquad w=1-u-v,
 \label{eq:last-entry-three-point-basis}
\end{equation}
reproducing the last-entry condition obeyed by the known three-point
remainder symbols~\cite{Dixon:2020bbt}.
At four points, a basis consists of the logarithmic differentials of
the following twelve ratios:
\begin{equation}
 \begin{aligned}
 &\frac{\mathcal L_{i,i-2}}{\mathcal L_{i,i+2}},
 \quad i=1,2,3,
 &\qquad
 &\frac{\mathcal L_{i,i+3}\mathcal L_{i+1,i-2}}
        {\mathcal L_{i,i+2}\mathcal L_{i+1,i+3}},
 \quad i=1,2,3,4,\\
 &\frac{\mathcal L_{i,i+4}\mathcal L_{i+2,i+6}}
        {\mathcal L_{i,i+2}\mathcal L_{i+2,i+4}},
 \quad i=1,2,
 &\qquad
 &\frac{\mathcal L_{i,i+3}\mathcal L_{i+1,i+4}
              \mathcal L_{i+2,i+6}}
        {\mathcal L_{i,i+2}\mathcal L_{i+1,i+3}
              \mathcal L_{i+2,i+4}},
 \quad i=1,2,3 .
 \end{aligned}
 \label{eq:last-entry-four-point-basis}
\end{equation}
All indices remain on the periodic cover, as in
\eqref{eq:last-entry-periodic-brackets}.
We compare this four-point candidate space with the complete two- and
three-loop remainder symbols
of ref.~\cite{Dixon:2022xqh}. Their last-entry spaces have dimensions nine
and twelve, respectively: the two-loop space is contained in our
candidate space, while the three-loop space coincides with it.

\section{Towards an all-loop derivation}
\label{sec:all-loop}

Within the periodic Wilson-loop description, the proposed equation requires
a relation between two insertions: the supersymmetry variation of the
superconnection and the collinear kink obtained by adding a cusp.
The amplitude descent construction identifies their local field content
\cite{Bullimore:2011kg,Caron-Huot:2011dec}. For the periodic contour, this
identification must also hold after summing the images, integrating along
an edge and making the finite BDS subtraction. We first describe the local
construction, then derive an analytic relation between the tree overlap
and the one-loop logarithmic source. This calculation identifies the
one-loop fermion energy correction entering the FFOPE; the finite matching
of the integrated insertions remains to be established.

\subsection{The supersymmetry anomaly and the collinear kink}

Let $\mathbb W_n$ be the periodic super Wilson-loop operator and
$W_n=\langle\mathbb W_n\rangle_{\rm per}^{[0]}$ its Grassmann-degree-zero
expectation value. The geometric generator \eqref{eq:qbar-fixed-q}
implements the supersymmetry of the self-dual theory. The correction to
the field transformation in the full theory contains
\cite{Bullimore:2011kg}
\begin{equation}
 \delta^{(1)}A=i|\psi^A\rangle[\bar\epsilon_A|,
 \qquad
 \delta^{(1)}|\bar\psi_A]
 =-\frac{i}{2}|\bar\epsilon_C]
 [\phi^{CB},\phi_{BA}] .
 \label{eq:field-variation}
\end{equation}
We suppress the common coupling factor in these field transformations.
The corresponding Ward identity has the form
\begin{equation}
 \qb W_n
 =-\langle\delta^{(1)}\mathbb W_n\rangle_{\rm per}^{[1]}+B_W .
 \label{eq:ward-identity}
\end{equation}
The superscript $[1]$ selects the term linear in the external fermionic
coordinates, as required for the variation of the MHV component.
The insertions in \eqref{eq:ward-identity} are dressed by the Wilson lines
along the rest of the contour. A periodic regulator may also contribute
through the variation of its boundary prescription; these terms are
denoted by $B_W$.

The scalar term can be treated directly in this projection. Define
$C^C{}_A=[\phi^{CB},\phi_{BA}]$ and let $\mathbb U_n(x)$ be the bosonic
holonomy based at its insertion point. An $SU(4)$-preserving regulator
implies
\begin{equation}
 \left\langle\operatorname{tr}
 [C^C{}_A(x)\mathbb U_n(x)]\right\rangle_{\rm per}
 =\kappa(x)\delta^C{}_A .
 \label{eq:scalar-insertion-singlet}
\end{equation}
The scalar duality relation
$\phi^{AB}=\tfrac12\epsilon^{ABCD}\phi_{CD}$ gives $C^A{}_A=0$,
so $\kappa(x)=0$. At Grassmann degree one, the scalar-commutator term
already carries one external $\theta$ and is therefore dressed only by
this bosonic holonomy. Its expectation vanishes by
\eqref{eq:scalar-insertion-singlet}. The fermionic term instead contracts
with the part of the superconnection linear in $\theta$. This is a
selection rule for the MHV expectation value, not a vanishing statement
for the scalar operator.

The same insertion arises locally from the extra cusp shown in
figure~\ref{fig:qbar-insertion}. Let $D_{\alpha A}$ be the gauge-covariant
fermionic derivative in chiral superspace, and let $\mathcal W_{AB}$
denote the fermionic supercurvature. Three derivatives with respect to
the insertion fermions produce its descendant
\begin{equation}
 \frac{\epsilon^{ABCD}}{3!}|D_B\mathcal W_{CD}\rangle
 =i|\psi^A\rangle+G|\theta^A\rangle
 +\frac12|\theta^B\rangle C^A{}_B+O(\theta^2),
 \label{eq:supercurvature-descendant}
\end{equation}
where $G_{\alpha\beta}$ is the auxiliary anti-self-dual two-form in the
self-dual formulation. The Grassmann measure also projects out the
$\chi_\star^4$ term of the inserted holomorphic frame. In the local
construction this is implemented by
\begin{equation}
 {\mathbb P^E}_A={\delta^E}_A
 -\frac14\theta^{E\alpha}
       \frac{\partial}{\partial\theta^{A\alpha}} .
 \label{eq:kink-projector}
\end{equation}
It removes the $G|\theta^A\rangle$ term; the remaining fermion and scalar
terms reproduce the field content of \eqref{eq:field-variation}
\cite{Bullimore:2011kg}. The selected $\mathrm N^{k+1}\mathrm{MHV}$ source
has Grassmann degree $4(k+1)$; three Grassmann integrations leave degree
$4k+1$, matching a $\qb$ variation. On the periodic contour the cusp
and its images must be inserted together. The local construction fixes
this operator structure, while its finite integrated normalization
requires the analysis below.

\subsection{The fermion kernel and finite normalization}
\label{sec:fermion-dispersion}

The FFOPE separates propagation along the flux tube from the
terminal transition that couples the state to the operator
\cite{Sever:2020jjx,Sever:2021xga}. These two ingredients enter the
insertion problem differently. Logarithms of the transverse collinear
parameter probe the energy of the propagating excitation. Finite
overlaps and endpoint terms also depend on its coupling to the terminal
transition. We can calculate the former directly from the three-point
source.

We use the three-point collinear calculation of
appendix~\ref{app:three-point-kernel}, with $u,v,w>0$, $u+v+w=1$,
and $\tau=z/(1-z)$. After stripping the common Grassmann factor and
dotted insertion vector, the bare tree insertion and the one-loop coefficient of
$\log\varepsilon$ are
\begin{equation}
 \begin{aligned}
 k_T(\tau)
 &=-\frac1{vw\,\tau}
   +\frac1{vs(\tau+u)}
   +\frac1{ws(\tau+u^{-1})},\qquad s=v+w,\\
 k_\ell(\tau)
 &=2\frac{\dd}{\dd\tau}
   \mathcal P_\ell\!\left(\frac{\tau}{1+\tau}\right).
 \end{aligned}
 \label{eq:tree-overlap-density}
\end{equation}
Here $\mathcal P_\ell(z)$ is the primitive in
\eqref{eq:three-point-log-primitive}; the factor of two follows from
$\log\rho=2\log\varepsilon+\log w+O(\varepsilon)$.
The tree insertion is evaluated before the subtraction in
\eqref{eq:mhv-equation}. Its normalization follows from
\eqref{eq:nmhv-tree-normalization} and \eqref{eq:three-point-box-kernels}.

To resolve the longitudinal modes, define
\begin{equation}
 \begin{split}
 I_0(\nu)&=\int_0^\infty\dd\tau\,\tau^\nu k_T(\tau),\\
 I_{1,\ell}(\nu)&=\int_0^\infty\dd\tau\,\tau^\nu k_\ell(\tau),
 \qquad 0<\operatorname{Re}\nu<1 .
 \end{split}
 \label{eq:mellin-convention}
\end{equation}
The tree transform is
\begin{equation}
 I_0(\nu)=-\frac{\pi}{\sin\pi\nu}
 \left(\frac{u^\nu}{vs}+\frac{u^{-\nu}}{ws}\right).
 \label{eq:tree-mellin-transform}
\end{equation}
Differentiated beta integrals give the exact identity
\begin{equation}
 \boxed{\ I_{1,\ell}(\nu)=h_\psi(\nu)I_0(\nu),\qquad
 h_\psi(\nu)=
 2\bigl[\psi(1+\nu)+\psi(1-\nu)-2\psi(1)\bigr]\ } .
 \label{eq:fermion-dispersion-test}
\end{equation}
Here $\psi(z)=\Gamma'(z)/\Gamma(z)$ is the digamma function.
The derivation is given in appendix~\ref{app:fermion-mellin}. It applies to
generic positive $u,v,w$ throughout the stated strip, including zeros of
$I_0$ where a ratio of the two transforms would be undefined.

The logarithmic transform actually converges in the larger strip
$-1<\operatorname{Re}\nu<1$. Although the tree transform has a pole at
$\nu=0$, the product in \eqref{eq:fermion-dispersion-test} is regular:
\begin{equation}
 I_0(\nu)=-\frac1{vw\,\nu}+O(1),\qquad
 h_\psi(\nu)=-4\zeta_3\nu^2+O(\nu^4),\qquad
 I_{1,\ell}(\nu)=\frac{4\zeta_3}{vw}\nu+O(\nu^2).
 \label{eq:fermion-zero-mode}
\end{equation}
Consequently
\begin{equation}
 \int_0^\infty\dd\tau\,k_\ell(\tau)=0,\qquad
 \int_0^\infty\dd\tau\,\log\tau\,k_\ell(\tau)
 =\frac{4\zeta_3}{vw}.
 \label{eq:fermion-log-moment}
\end{equation}
The first equality follows also from \eqref{eq:three-point-log-cancellation};
the second resolves the first nonvanishing longitudinal moment.

With the OPE momentum convention $\nu=ip/2$, the multiplier gives
\begin{equation}
 E_\psi(p)-1=g^2h_\psi(ip/2)+O(g^4),
 \label{eq:fermion-energy}
\end{equation}
the one-loop large-fermion dispersion relation in the normalization
$\Gamma_{\rm cusp}=4g^2+O(g^4)$
\cite{Caron-Huot:2011dec,Belitsky:2015kda}.
On this line the logarithmic transform converges, while the bare tree
transform is understood by analytic continuation from its convergence
strip; its pole at $p=0$ is retained.
The analytically continued one-loop multiplier vanishes at $p=0$,
consistently with the protected supersymmetry mode of the amplitude OPE
\cite{Basso:2014hfa}. Here this vanishing follows directly from the
projected form-factor source. Identifying the finite-coupling mode also
requires the appropriate continuation between fermion sheets.

The finite matching problem remains. Let $\boldsymbol\Psi_i(\sigma)$
with $\sigma=\log\tau$ be the Wilson-line-dressed Ward insertion, normalized so that its
contribution to \eqref{eq:ward-identity} carries the overall factor $g^2$,
with the sign absorbed into $\boldsymbol\Psi_i$.
Let $\mathbf K_i(\sigma;\varepsilon)$ be the kink insertion with its
Grassmann projection. A possible renormalized relation is
\begin{equation}
 \mathbf K_i(\sigma;\varepsilon)
 =\int\dd\sigma'\,
 F_g(\sigma,\sigma';\varepsilon)\boldsymbol\Psi_i(\sigma')
 +\mathbf E_i(\sigma;\varepsilon),
 \label{eq:integrated-matching}
\end{equation}
where $\mathbf E_i$ contains contact and endpoint contributions. The
logarithmic evolution measured by $h_\psi$ does not determine the finite
kernel $F_g$. In a translation-invariant bulk frame, a sufficient
condition for its bulk normalization is
\begin{equation}
 \int\dd\sigma\,F_g(\sigma,\sigma';\varepsilon)=F_0(g),
 \label{eq:left-zero-mode}
\end{equation}
independent of $\sigma'$. The amplitude equation supplies the calibration
\begin{equation}
 \frac{g^2}{F_0(g)}=\frac{\Gamma_{\rm cusp}(g)}4,\qquad
 F_0(g)=1+2\zeta_2g^2+O(g^4).
 \label{eq:cusp-calibration}
\end{equation}
For the form-factor equation, this normalization must survive the
periodic image sum and the finite subtraction.

At the seam, an adjacent segment is a translated image of the contour.
It must participate in the same soft and collinear subtraction as an
ordinary neighbor. The two terms in \eqref{eq:mhv-equation} have cancelling
endpoint residues when regulated together, but transporting the cutoff
between coordinates can leave a finite contribution. The explicit
$C_{\rm end}$ in \eqref{eq:three-point-endpoint-correction} is an example
already encountered in the three-point calculation. Such contributions,
and the regulated Ward term $B_W$, have to be included when matching
\eqref{eq:integrated-matching} after integration.

Finally, the bosonic duality in a common scheme gives
\begin{equation}
 W_n=\frac{F_{n,0}}{F_{n,0}^{\rm tree}}=D_ne^{R_n}
 \label{eq:wilson-form-factor}
\end{equation}
\cite{Brandhuber:2010ad,Basso:2023bwv,Li:2024rkq}.
Dividing the matched insertion by $W_n$ and removing $\qb\log D_n$
would then yield $\qb R_n$. Before the BDS factors are removed, the
NMHV insertion divided by the MHV loop carries
$(D_{n+1}/D_n)e^{R_{n+1}-R_n}\mathcal R_{n+1,1}$.
The adjacent-multiplicity ratio $D_{n+1}/D_n$, including its finite
regulator-dependent contribution, must be absorbed in this matching.
The BDS-normalized insertion then has the exponential ratio appearing
in \eqref{eq:mhv-equation}, with the BDS reference producing the tree
subtraction.
An all-loop derivation therefore requires the finite integrated
Ward--kink matching, including its terminal and endpoint terms, in one
periodic regulator. The calculation above identifies the one-loop
logarithmic evolution of the form-factor source with the fermion
dispersion relation.

\section{Discussion and outlook}
\label{sec:discussion}

The periodic $\bar q$ equation relates derivatives of an $n$-point
$\mathrm N^k\mathrm{MHV}$ form factor to a collinear integral of
lower-loop $\mathrm N^{k+1}\mathrm{MHV}$ data at $n+1$ points.  The analytic one-loop NMHV test and the reconstruction of
the complete three-point two-loop MHV remainder give complementary evidence
for this relation, supplemented by the higher-point MHV comparisons.
For a parity-even MHV remainder, the parity-related equation supplies the
missing derivative directions, allowing the full differential to be
reconstructed from lower-loop sources.  At higher multiplicity this
requires compatible, closed source one-forms. 

A particularly useful next check is the three-loop MHV equation,
which is sensitive to the finite normalization.  Expanding
\eqref{eq:mhv-equation} gives
\begin{equation}
 \qb R_n^{(3)}=\sum_i\cI_i\left[
 \cR_{n+1,1}^{(2)}
 +(R_{n+1}^{(2)}-R_n^{(2)})\cR_{n+1,1}^{(0)}
 -2\zeta_2\cR_{n+1,1}^{(1)}\right].
 \label{eq:three-loop-prediction}
\end{equation}
This is the first order sensitive both to the nontrivial exponential
conversion between BDS normalizations and to a higher coefficient of the
cusp anomalous dimension.  The known four-point two-loop NMHV
symbol~\cite{He:2026ddp} provides part of the input for the three-point
specialization.  A complete test also needs beyond-the-symbol
contributions, such as $\zeta_2$ times a lower-weight function.
Tests at higher multiplicity would probe the equation in a larger
kinematic space and further test the compatibility of its parity-related
derivatives.

At three points, \eqref{eq:mhv-equation} would give an all-loop relation
between the MHV differential and a collinear integral of the four-point
NMHV form factor, with the specified BDS conversion and tree subtraction.
The three-point MHV form factor is already known through eight
loops~\cite{Dixon:2022rse}.  These data can therefore also be used in the
opposite direction, to constrain integrated combinations of higher-loop
four-point NMHV data before the full four-point function is known.
This offers a way to connect the three- and four-point bootstraps across
helicity sectors and loop orders.

The MHV last-entry candidates of section~\ref{sec:mhv-last-entries}
recover the known three-point condition and accommodate the last entries
of the known four-point remainder symbols, under the assumptions stated
there. Further bootstrap
constraints would come from relations among the corresponding coefficient
functions and from extending the analysis to the NMHV sector, as for
amplitudes~\cite{Caron-Huot:2011dec,He:2020vob}.
There the last entries must be studied together with
the rational or algebraic prefactors and their Grassmann structure:
the $\bar q$ action on the full superfunction and identities among
prefactors determine which combinations of coefficient functions are
constrained.  The four-point NMHV bootstrap~\cite{He:2026ddp} provides
a concrete setting in which to extract and use these relations.

A related question is whether iterating the equation can explain the
multiple-final-entry relations observed in the three-point
bootstrap~\cite{Dixon:2022rse}.
Combined with the alphabets and analytic structure studied in
refs.~\cite{He:2026kdq,He:2026ddp}, such relations could substantially
reduce the ansatz spaces for three- and four-point form factors.
The FFOPE supplies complementary near-collinear expansion
coefficients and boundary data~\cite{Sever:2020jjx,Sever:2021xga}, which
could then fix the remaining bootstrap parameters.

The relation between three-point MHV and four-point NMHV form factors
may also help explain the simplicity of their function spaces and the
antipodal structures found in recent work.  The conjectured antipodal
duality relates the three-point MHV form factor to the six-point MHV
amplitude on its parity-preserving surface~\cite{Dixon:2021tdw}, while
an antipodal self-duality has been observed for
the four-point MHV form factor on its parity-preserving
surface~\cite{Dixon:2022xqh,Dixon:2024yvq}.
The cluster-algebra description of the antipodal map~\cite{He:2026kdq}
provides a kinematic setting for this comparison, while the four-point
NMHV bootstrap~\cite{He:2026ddp} provides data beyond MHV.
Since the antipode reverses symbol entries, understanding
the origin of last-entry relations could shed light on their correspondence
with first-entry constraints under these dualities.

The FFOPE offers a route to extending the matching discussed
in section~\ref{sec:all-loop} to finite coupling.  In the amplitude
OPE, the $\bar Q$ insertion is represented by a zero-momentum fermion and
charged pentagon transitions~\cite{Basso:2014hfa,Basso:2015rta}.  The periodic
problem requires a charged terminal form-factor transition with the same
finite subtraction and endpoint prescription; the singlet selection rule
for the ordinary terminal transition does not determine it.  Identifying
this charged transition with the collinear insertion would connect the
periodic Ward identity to the finite-coupling flux-tube description
\cite{Sever:2021xga,Basso:2023bwv}.  Such a construction could lead to
nonperturbative relations between helicity sectors, beyond their use as
constraints on perturbative bootstraps.  Extending the equation away from
zero fermionic monodromy and reconstructing complete differentials at
higher helicity are further directions.

\acknowledgments

We thank Chi Zhang for helpful comments on the draft.
OpenAI Codex was used to assist with literature searches, exploratory
calculations and language editing. The authors reviewed and revised all
AI-assisted material and take full responsibility for the content. The work
of S.H. is supported by the National Natural Science Foundation of China under
Grant Nos.~12225510 and 12247103, and by the New Cornerstone Science
Foundation.

\appendix
\section{Monodromy, normalization, and insertion conventions}
\label{app:monodromy}

With the spinor conventions of section~\ref{sec:equation}, let
$Z_i=(\lambda_i,\mu_i)$ and define the translation block $Q$ by its
action $(Q\lambda_i)^{\dot\alpha}
=q^{\alpha\dot\alpha}\lambda_{i\alpha}$.
The bosonic monodromy is represented by
\begin{equation}
 P=
 \begin{pmatrix}
 \mathbf1_2&0\\ Q&\mathbf1_2
 \end{pmatrix},
 \qquad Z_{i+n}=PZ_i .
 \label{eq:monodromy-matrix}
\end{equation}
For transformations of the full supertwistor $\mathcal Z_i=(Z_i,\chi_i)$,
we extend $P$ to act trivially on $\chi_i$ at $\gamma=0$.
A linear transformation $\mathcal Z_i'=G\mathcal Z_i$ induces
$P'=GPG^{-1}$.  Writing $G=1+T$, with $T$ including the infinitesimal
transformation parameters, gives $\delta P=[T,P]$.
For each $R$-symmetry index $A$, the dotted components of $\bar Q_a^A$
used in \eqref{eq:qbar-fixed-q} preserve $P$ on the $\gamma=0$ slice.
The undotted components instead act as
\begin{equation}
 \delta\lambda_i^\alpha=\bar\epsilon_A^\alpha\chi_i^A,
 \qquad \delta\mu_i^{\dot\alpha}=0 .
 \label{eq:undotted-qbar-action}
\end{equation}
At fixed $q$, the variation of the original periodicity condition is
\begin{equation}
 \delta\bigl(
 \mu_{i+n}^{\dot\alpha}-\mu_i^{\dot\alpha}
 -q^{\beta\dot\alpha}\lambda_{i\beta}\bigr)
 =-q^{\beta\dot\alpha}\bar\epsilon_{\beta A}\chi_i^A .
 \label{eq:undotted-monodromy-variation}
\end{equation}
Thus the undotted transformation does not preserve the original
fixed-$q$ periodicity.  Allowing the monodromy to transform by conjugation
instead generates an additional $\mu$--$\chi$ block, outside the chosen
pure-translation form.  Labels outside a fundamental period are kept on
the universal cover;
when needed, they are represented by their images under $P$ rather than
identified with labels in the fundamental period.

In the spinor frame used for the analytic and numerical comparisons, we set
$c_i=\langle i-1\,i\rangle/\langle i\,i+1\rangle$.
The labels are understood on the periodic cover.
The nonzero coefficient $c_i'$ in
\eqref{eq:collinear-curve} specifies the transverse approach at order
$\varepsilon^2$.  For real kinematics, its sign selects the side from
which the collinear configuration is approached.  This choice must be
specified together with the analytic continuation of the scalar functions
and used consistently in the source and subtraction terms.
A change of the longitudinal variable leaves the integral invariant when
the endpoint cutoff is transformed consistently.  Applying separate
changes of variable to endpoint-divergent terms while keeping their
numerical cutoffs fixed can shift their finite parts.  We therefore use
the same collinear parametrization and a common cutoff for all source and
subtraction terms before taking the endpoint limit.

We use dimensional regularization in $D=4-2\epsilon$, with the loop
normalization of ref.~\cite{Brandhuber:2012vm} and the coupling convention
in \eqref{eq:coupling-convention}. Dependence on the common regularization
scale is suppressed. For the dimensionally regulated MHV ratio
\begin{equation}
 \mathcal M_n(\epsilon)
 =\frac{F_{n,0}}{F_{n,0}^{\rm tree}}
 =1+\sum_{L\geq1}g^{2L}\mathcal M_n^{(L)}(\epsilon),
\end{equation}
the BDS factor is
\begin{equation}
 \log D_n=\sum_{\ell\geq1}g^{2\ell}
 \left[f^{(\ell)}(\epsilon)
 \mathcal M_n^{(1)}(\ell\epsilon)+C^{(\ell)}\right]
 +\mathcal O(\epsilon),
 \qquad
 f^{(\ell)}(\epsilon)=f_0^{(\ell)}+\epsilon f_1^{(\ell)}
 +\epsilon^2f_2^{(\ell)},
 \label{eq:bds-factor}
\end{equation}
with $f_0^{(\ell)}=\Gamma_{\rm cusp}^{(\ell)}/4$ and
$\Gamma_{\rm cusp}(g)=\sum_{\ell\geq1}g^{2\ell}
\Gamma_{\rm cusp}^{(\ell)}$.  The coefficient $f_1^{(\ell)}$ is fixed
by the collinear anomalous dimension, while $f_2^{(\ell)}$ and
$C^{(\ell)}$ are kinematics-independent constants specifying the finite
BDS subtraction~\cite{Bern:2005iz,Brandhuber:2012vm}.
We follow the form-factor BDS convention of ref.~\cite{Brandhuber:2012vm},
in which $R_2^{(2)}=0$.
The same regulator and
analytic continuation are used at multiplicities $n$ and $n+1$.  We choose
the logarithmic remainder $R_n$ to vanish at tree level and one loop.

\section{Details of the perturbative calculations}
\label{app:comparison-details}

We collect the analytic details of the one-loop NMHV comparison and the
three-point two-loop MHV reconstruction in section~\ref{sec:evidence}.

\subsection{Details of the one-loop NMHV comparison}
\label{app:one-loop-nmhv-comparison}

We specify the integrated one-loop input and the BCFW tree expressions
used in \eqref{eq:one-loop-nmhv}, followed by the component projection and
analytic comparison.  All twistor labels below lie on the periodic cover.
For an ordered tuple of $n$ supertwistors, its images obey
$\mathcal Z_{i+n}=P\mathcal Z_i$.

\paragraph{Integrated one-loop input.}
We use the ordered four- and five-brackets
\begin{equation}
 \begin{split}
 \langle abcd\rangle&=\det(Z_a,Z_b,Z_c,Z_d),\\
 [abcde]&=\frac{\delta^{0|4}\!\left(
 \chi_a\langle bcde\rangle+\chi_b\langle cdea\rangle
 +\chi_c\langle deab\rangle+\chi_d\langle eabc\rangle
 +\chi_e\langle abcd\rangle\right)}
 {\langle abcd\rangle\langle bcde\rangle\langle cdea\rangle
 \langle deab\rangle\langle eabc\rangle} .
 \end{split}
 \label{eq:nmhv-ordered-brackets}
\end{equation}
Here $\delta^{0|4}(L)=L^1L^2L^3L^4$.  In this ordering, the box coefficients
in \eqref{eq:four-point-nmhv-input} are fixed by
\begin{align}
 c_1^{1\mathrm m}={}&[2,3,4,0,1]
 -\frac{\langle1,-1,0,4\rangle\langle1,3,4,-1\rangle}
 {\langle5,3,4,1\rangle\langle4,0,3,-1\rangle}[-1,0,3,4,1],
 \label{eq:nmhv-box-coefficient}\\
 c_i^{1\mathrm m}={}&\left.c_1^{1\mathrm m}\right|_{j\mapsto j+i-1},
 \qquad \cR_{4,1}^{(0)}=\frac12\sum_{i=1}^4c_i^{1\mathrm m} .
 \label{eq:nmhv-tree-normalization}
\end{align}
The shift acts on every integer label, including its periodic images.
These are the ordered five-bracket representatives of the coefficients
in refs.~\cite{Bork:2012tt,Bork:2014eqa}.

The finite box functions are
\cite{Bork:2012tt,Bianchi:2018rrj}
\begin{align}
 F^{1\mathrm m}(s,t;M)={}&
 \Li_2(1-M/s)+\Li_2(1-M/t)+\tfrac12\log^2(s/t)+\zeta_2,
 \nonumber\\
 F^{2\mathrm{mh}}(s,t;M,N)={}&
 \Li_2(1-M/t)+\Li_2(1-N/t)+\tfrac12\log^2(s/t)
 \nonumber\\
 &-\tfrac12\log(s/M)\log(s/N),\nonumber\\
 F^{2\mathrm{me}}(s,t;M,N)={}&
 \Li_2(1-M/s)+\Li_2(1-M/t)
 +\Li_2(1-N/s)+\Li_2(1-N/t)\nonumber\\
 &-\Li_2(1-MN/(st))+\tfrac12\log^2(s/t).
 \label{eq:nmhv-finite-boxes}
\end{align}
Finite parts are obtained by removing the unexpanded infrared power
terms from the reduced integrals, with the same convention in both
helicity sectors.  The finite MHV correction in that convention is
\cite{Brandhuber:2010ad,Bianchi:2018rrj}
\begin{equation}
 \mathcal M_{4,\mathrm{fin}}^{(1)}=
 \sum_{i=1}^4\left[
 F^{1\mathrm m}(u_i,u_{i+1};v_i)
 +F^{2\mathrm{me}}(v_{i+1},v_{i+2};1,u_{i+2})\right].
 \label{eq:nmhv-mhv-subtraction}
\end{equation}

For the three-mass triangles define
\begin{equation}
 y^2=(1+U-V)^2-4U,\qquad h_\pm=\frac{1+U-V\pm y}{2} .
 \label{eq:nmhv-triangle-roots}
\end{equation}
The three-mass triangle function is~\cite{Bianchi:2018rrj}
\begin{equation}
 F^{3\mathrm m}(U,V)
 =\Li_2(h_+)-\Li_2(h_-)
 +\frac12\log U\log\frac{1-h_+}{1-h_-},
 \label{eq:triangle-function}
\end{equation}
We use the cross ratios $(U,V)$ as arguments; the root variables in
that reference are $(z,\bar z)=(h_+,h_-)$.
Label the two triangle channels by $i=1,2$ and set
\begin{equation}
 U_i=u_{i+2},\qquad V_i=u_i,\qquad
 K_{i,h}=hq-(x_{i+4}-x_{i+2}) .
 \label{eq:nmhv-triangle-channels}
\end{equation}
For each channel, $K_{i,h}$ is null at either root
$h=h_\pm(U_i,V_i)$.  Let $\lambda_h$ denote its spinor, with the
channel label suppressed.  The coefficients, including the inverse
square root, are~\cite{Bork:2012tt,Bianchi:2018rrj}
\begin{equation}
 c_i^{3\mathrm m}=
 \sum_{h=h_+,h_-}
 \frac{\langle i+3,i+4\rangle\langle i+1,i+2\rangle
 \delta^{0|4}\!\left((\theta_{i+2}^{\alpha A}-\theta_{i+4}^{\alpha A})\lambda_{h\alpha}\right)}
 {q^2y(U_i,V_i)h(1-h)
 \langle i+4,h\rangle\langle i+3,h\rangle
 \langle i+2,h\rangle\langle i+1,h\rangle} .
 \label{eq:nmhv-triangle-coefficients}
\end{equation}

\paragraph{BCFW tree input.}
We use the periodic momentum-twistor BCFW recursion of
ref.~\cite{Bork:2014eqa} to construct the three tree ratios entering
the right-hand side of \eqref{eq:one-loop-nmhv}.
Define
\begin{equation}
 \mathcal B_n=
 -\frac{\langle1,2,3,2-n\rangle\langle1,2-n,3-n,3\rangle}
 {\langle1,2,3,1+n\rangle\langle2,3,2-n,3-n\rangle}
 [1,2,3,2-n,3-n] .
 \label{eq:nmhv-bcfw-minimal}
\end{equation}
Starting with $\cR_{3,1}^{(0)}=\mathcal B_3$, the NMHV ratios are
\begin{align}
 \cR_{n,1}^{(0)}(1,\ldots,n)={}&
 \cR_{n-1,1}^{(0)}(1,3,\ldots,n)
 +\sum_{j=4}^{n}[1,2,3,j,j+1]\nonumber\\
 &+\sum_{j=3}^{n-1}[1,2,3,j-n,j+1-n]+\mathcal B_n,
 \qquad n=4,5 .
 \label{eq:nmhv-bcfw-trees}
\end{align}
Explicit arguments of $\cR_{n,k}^{(0)}$ or $\mathcal B_n$ specify an
ordered configuration.  Its entries are relabelled consecutively, and
its periodic images are generated with period $n$ and the same physical
translation $P$.

Define the intersection supertwistors
\begin{align}
 I_j&=(j,j+1)\cap(1,2,3),\qquad
 H_j=(2,3)\cap(1,j,j+1),\nonumber\\
 (a,b)\cap(c,d,e)&=
 \mathcal Z_a\langle bcde\rangle-\mathcal Z_b\langle acde\rangle .
 \label{eq:nmhv-bcfw-intersections}
\end{align}
The N$^2$MHV ratios are
\begin{align}
 \cR_{4,2}^{(0)}(1,2,3,4)={}&
 [1,2,3,-1,0]\mathcal B_3(I_{-1},H_{-1},3),
 \label{eq:nmhv-bcfw-four-n2}\\
 \cR_{5,2}^{(0)}(1,2,3,4,5)={}&
 \cR_{4,2}^{(0)}(1,3,4,5)
 +[1,2,3,4,5]\mathcal B_3(1,I_4,5)\nonumber\\
 &+[1,2,3,5,6][I_5,H_5,3,4,5]\nonumber\\
 &+[1,2,3,-2,-1]\mathcal B_3(I_{-2},H_{-2},3)\nonumber\\
 &+[1,2,3,-1,0]\cR_{4,1}^{(0)}(I_{-1},H_{-1},3,4)\nonumber\\
 &+\mathcal B_5[I_{-3},-2,-1,0,1] .
 \label{eq:nmhv-bcfw-five-n2}
\end{align}
Intersections are evaluated on the corresponding ordered configuration.
In particular, $\cR_{4,2}^{(0)}(1,3,4,5)$ in
\eqref{eq:nmhv-bcfw-five-n2} uses the four-point periodic cover.

\paragraph{Coordinates and dotted components.}
For the analytic comparison we use the generic patch
\begin{equation}
 \begin{gathered}
 Z_1=(1,0,0,0),\quad Z_2=(0,1,0,0),\quad
 Z_3=(1,1,m_3,n_3),\quad Z_4=(1,\kappa,m_4,n_4),\\
 Q=\mathbf1_2,\qquad
 \chi_1=\chi_2=0,\qquad \chi_3=\alpha,\quad\chi_4=\beta .
 \end{gathered}
 \label{eq:four-point-generic-patch}
\end{equation}
Here $Q$ is the monodromy block of appendix~\ref{app:monodromy}, so
$q^2=1$ in this frame, and $(\kappa,m_3,n_3,m_4,n_4)$ are five
independent coordinates.  We choose the fermionic region origin so that
$\chi_1=\chi_2=0$, which is possible for $\langle12\rangle\ne0$.
To compare the two sides component by component, define the Grassmann bases by
\begin{equation}
 \prod_{A=1}^4(\alpha^A+t\beta^A)=\sum_{r=0}^4t^r\Phi_r,
 \qquad
 \prod_{A=2}^4(\alpha^A+t\beta^A)=\sum_{r=0}^3t^r\Psi_r .
 \label{eq:nmhv-grassmann-basis}
\end{equation}
Here $t$ is a commuting auxiliary variable, and products are ordered by
increasing $A$.  We write $\cR_{4,1}^{(1)}=\sum_{r=0}^4f_r\Phi_r$,
where $f_r$ are bosonic coefficient functions.
The dotted variation for free index $A=1$ is
\begin{equation}
 \qb^1_{\dot a}\cR_{4,1}^{(1)}
 =\alpha^1\beta^1\sum_{r=0}^3J_{\dot a}^{(r)}\Psi_r,\qquad
 J_{\dot a}^{(r)}=
 \partial_{\mu_3^{\dot a}}f_{r+1}
 -\partial_{\mu_4^{\dot a}}f_r .
 \label{eq:four-point-components}
\end{equation}
The other free $R$-symmetry indices follow by relabelling.  It therefore
suffices to compare the eight components $J_{\dot a}^{(r)}$, with
$r=0,\ldots,3$ and $\dot a=\dot1,\dot2$.
The derivatives are taken at fixed $Q$ and
fixed undotted spinors $\lambda_i$.

Direct differentiation of the coefficients in
\eqref{eq:nmhv-box-coefficient}--\eqref{eq:nmhv-tree-normalization}
and \eqref{eq:nmhv-triangle-coefficients} gives
\begin{align}
 \qb c_i^{1\mathrm m}&=0,\qquad i=1,\ldots,4,\nonumber\\
 \qb c_i^{3\mathrm m}&=0,\qquad i=1,2.
 \label{eq:nmhv-coefficient-closure}
\end{align}
Since $\cR_{4,1}^{(0)}=\tfrac12\sum_i c_i^{1\mathrm m}$,
$\qb$ acts only on the box and triangle functions and the finite MHV
correction in \eqref{eq:four-point-nmhv-input}.
For the triangle function,
$h_+h_-=U$ and $(1-h_+)(1-h_-)=V$ give
\begin{align}
 2y\,\dd F^{3\mathrm m}={}&
 \left(2\log U+\frac{1-U-V}{U}\log V\right)\dd U\nonumber\\
 &+\left(\frac{1-U-V}{V}\log U+2\log V\right)\dd V .
 \label{eq:triangle-differential}
\end{align}
Together with the derivatives of the box functions in
\eqref{eq:nmhv-finite-boxes}, this expresses the left-hand components
$J_{\dot a}^{(r)}$ as sums of rational functions multiplying
$\log u_i$ and $\log v_i$.

\paragraph{Projection, integration and matching.}
We evaluate the right-hand side using the BCFW expressions above and
\eqref{eq:collinear-curve}, with $c'_e=1$ and $c_e$ as defined in
appendix~\ref{app:monodromy}.  On each edge, we perform the three
Grassmann integrations on the full source, then restrict the remaining
fermion to the insertion curve and take the collinear residue.
All periodic images share the same independent insertion fermion.
After stripping the dotted measure factor, the coefficients of
$\alpha^1\beta^1\Psi_r$ give the rational kernels $k_e^{(r)}(\tau)$,
with $e=1,\ldots,4$ and $r=0,\ldots,3$.

Denote the four entries of \eqref{eq:four-point-poles}, in the order
shown, by $a_{e,1},\ldots,a_{e,4}$.  The kernels have the decomposition
\begin{equation}
 k_e^{(r)}(\tau)=\sum_{j=1}^4\frac{b_{e,j}^{(r)}}{\tau+a_{e,j}},
 \qquad \sum_{j=1}^4b_{e,j}^{(r)}=0,
 \label{eq:nmhv-projected-kernel}
\end{equation}
where $b_{e,j}^{(r)}$ are the residues at $\tau=-a_{e,j}$.
Using \eqref{eq:logarithmic-integral}, restoring the dotted measure
factor in \eqref{eq:insertion} and summing the four edges gives
\begin{align}
 J_{\dot a,\mathrm{RHS}}^{(r)}={}&\sum_{e=1}^4 c_e(e-1,e,e+1)_{\dot a}
 \left[b_{e,1}^{(r)}\log\frac{v_{e+2}}{u_{e+2}}
 +b_{e,2}^{(r)}\log\frac{u_{e+1}}{v_e}
 +(b_{e,4}^{(r)}-b_{e,3}^{(r)})\log v_{e+1}\right].
 \label{eq:nmhv-integrated-components}
\end{align}
After collecting the coefficients of $\log u_i$ and $\log v_i$, we find
that all 64 rational coefficients agree identically with those obtained by
differentiating \eqref{eq:four-point-nmhv-input}.
The comparison is made on a common sheet with positive $u_i,v_i$,
with $0<h_-<h_+<1$ fixing the initial triangle branches in both channels.
Thus $J_{\dot a,\mathrm{RHS}}^{(r)}=J_{\dot a}^{(r)}$ on this patch;
the equality extends to other regions by consistent analytic continuation.

\subsection{Three-point two-loop MHV reconstruction}
\label{app:three-point-kernel}

We evaluate the single-edge function $K(u,v,w)$ in
\eqref{eq:one-edge-three-point-form} and use its cyclic sum to reconstruct
the three-point two-loop remainder.
The input is the four-point one-loop NMHV ratio
\eqref{eq:four-point-nmhv-input}.  Throughout, $u+v+w=1$ with
$u,v,w>0$, and the functions are evaluated on their real branches.

\paragraph{Collinear input.}
We use the three-point twistor configuration
\begin{equation}
 Z_1=(1,0,0,0),\quad Z_2=(0,1,0,0),\quad
 Z_3=(1,1,u+v,v),\quad
 P=\begin{pmatrix}\mathbf1_2&0\\\mathbf1_2&\mathbf1_2\end{pmatrix}.
 \label{eq:mellin-twistor-patch}
\end{equation}
The translation block defined in appendix~\ref{app:monodromy} is
$Q=\mathbf1_2$.  We choose $c_i'=+1$ on every edge.  For the third
edge, the inserted twistor is
\begin{equation}
 Z_4=Z_3-\varepsilon Z_2+\varepsilon\tau PZ_1+\varepsilon^2PZ_2.
 \label{eq:mellin-insertion-patch}
\end{equation}
The new fundamental period is $(Z_1,Z_2,Z_3,Z_4)$, with images
$Z_{i+4}=PZ_i$.
Set $s=1-u=v+w$ and
$z=\tau/(1+\tau)$, so that $0<z<1$.  At fixed $z$, the leading
collinear values of the four-point invariants
\eqref{eq:four-point-invariants} are
\begin{equation}
 (u_1,u_2,u_3,u_4)\simeq(u,zv,\rho,(1-z)w),\qquad
 (v_1,v_2,v_3,v_4)\simeq(u+sz,v,w,1-sz).
 \label{eq:three-point-pullback}
\end{equation}
Here $\simeq$ denotes the leading term of each component as
$\varepsilon\to0$.  We retain the small invariant
$\rho=u_3=w\varepsilon^2+O(\varepsilon^3)$, since its logarithm
$\log\rho=2\log\varepsilon+\log w+O(\varepsilon)$ contributes to the
longitudinal integrand.

The sum $F_i$ of the three finite box functions multiplying
$c_i^{1\mathrm m}$ in \eqref{eq:four-point-nmhv-input} can be written,
using \eqref{eq:nmhv-finite-boxes}, as
\begin{align}
 F_i={}&2\Li_2(1-1/v_i)+\zeta_2-L_iL_{i+1}
 +\frac12L_iL_{i+2}+\frac12L_{i+1}L_{i+3}\nonumber\\
 &+\log v_i\,(L_i+L_{i+1}-L_{i+2}-L_{i+3}),
 \label{eq:three-point-Fi}
\end{align}
where $L_i=\log u_i$ and the scalar indices are cyclic modulo four.
Let $F_i^0$ denote the term independent of $\log\rho$ after the leading
substitution \eqref{eq:three-point-pullback}.  The corresponding constant term of the
MHV correction $\mathcal M_{4,\mathrm{fin}}^{(1)}$ in
\eqref{eq:nmhv-mhv-subtraction} is
$M_4^0=F_2^0+F_3^0-2\Li_2(s)$.
The Grassmann integration and collinear residue of the coefficients
$c_i^{1\mathrm m}$ in
\eqref{eq:nmhv-box-coefficient}--\eqref{eq:nmhv-tree-normalization},
with the common dotted factor stripped off, give
\begin{align}
 K_1(z)&=\frac1{vs}\left[-\frac1z+\frac2{z+u/s}-\frac1{z-1}\right],
 \nonumber\\
 K_2(z)&=\frac1{vw}\left[-\frac1z+\frac1{z-1}\right],\nonumber\\
 K_4(z)&=\frac1{ws}\left[-\frac1z+\frac2{z-1/s}-\frac1{z-1}\right].
 \label{eq:three-point-box-kernels}
\end{align}
The common factor is $\chi_3^A(v,w)$ for the components
$(\dot1,\dot2)$.
These kernels include the Jacobian $\dd\tau/\dd z$;
$c_3^{1\mathrm m}$ does not contribute to the residue.
Of the two triangle channels $(u_3,u_1)$ and $(u_4,u_2)$ in
\eqref{eq:nmhv-triangle-coefficients},
the former degenerates, but the contribution of the root
that tends to zero survives the residue.  The discriminant of the other
channel remains nonzero; we denote its positive square root by $\Delta(z)$:
\begin{equation}
 \Delta(z)^2=\bigl(1-zv-(1-z)w\bigr)^2-4zv(1-z)w,
 \qquad \Delta(z)>0 .
 \label{eq:three-point-conic}
\end{equation}

\paragraph{The finite integrand.}
After the collinear projection, the leading integrand has the form
$\mathcal D_\ell(z)\log\rho+\mathcal D(z)$.
The logarithmic coefficient does not vanish pointwise.  Defining
\begin{equation}
 W_1=\log\frac{(u+sz)^2}{z(1-z)},\qquad
 W_2=\log\frac{z}{1-z},\qquad
 W_4=\log\frac{(1-sz)^2}{z(1-z)},
\end{equation}
we find $\mathcal D_\ell(z)=\partial_z\mathcal P_\ell(z)$, where
\begin{align}
 \mathcal P_\ell(z)={}&
 \frac{2\log u\,W_1-W_1^2}{4vs}
 +\frac{W_2^2-2\log u\,W_2}{4vw}\nonumber\\
 &+\frac{2\log u\,W_4-W_4^2}{4ws}
 +\frac{\log u\,W_2}{ws}.
 \label{eq:three-point-log-primitive}
\end{align}
Applying the common endpoint prescription of
appendix~\ref{app:monodromy} with $\delta\leq z\leq1-\delta$ gives
\begin{equation}
 \lim_{\delta\to0^+}
 [\mathcal P_\ell(1-\delta)-\mathcal P_\ell(\delta)]=0.
 \label{eq:three-point-log-cancellation}
\end{equation}
Thus the terms proportional to $2\log\varepsilon$ and to the accompanying
$\log w$ cancel after integration.  The remaining integrand determines $K$:
\begin{align}
 \mathcal D(z)={}&
 \sum_{i=1,2,4}K_i(z)\left(F_i^0-\frac12M_4^0\right)
 +\frac{2\Li_2(s)}{ws\,z(1-z)}\nonumber\\
 &-2\frac{(1+u)z-(1-w)}{vw\,z(1-z)}
 \frac{F^{3\mathrm m}(zv,(1-z)w)}{\Delta(z)},\nonumber\\
 K(u,v,w)={}&\lim_{\delta\to0^+}
 \int_\delta^{1-\delta}\dd z\,\mathcal D(z).
 \label{eq:three-point-source-density}
\end{align}
Here $F^{3\mathrm m}$ is the three-mass triangle function
defined in \eqref{eq:triangle-function}.
The first term contains the box
contributions and MHV subtraction; the second comes from the degenerate
triangle, and the last from the other triangle channel.  In the last term
we used $F^{3\mathrm m}(U,V)=F^{3\mathrm m}(V,U)$ to order the arguments.
The endpoint singularities cancel in their sum, so all terms must use
the same cutoff.

\paragraph{Analytic longitudinal integration.}
Let $\mathcal D_B(z)$ denote the sum of the first two terms of
$\mathcal D(z)$ in \eqref{eq:three-point-source-density}, containing
the boxes and the degenerate triangle.  We integrate these terms in $z$.
For the remaining triangle, rationalize the root by
\begin{equation}
 z=\frac{t(1-w+wt)}{u+st},\qquad
 1-z=\frac{(1-t)(u+wt)}{u+st},\qquad
 \frac{\dd z}{\Delta(z)}=\frac{\dd t}{u+st}.
 \label{eq:three-point-rationalization}
\end{equation}
This maps $0<z<1$ monotonically to $0<t<1$.  Including the Jacobian,
the last term of \eqref{eq:three-point-source-density} becomes
$\mathcal D_T(t)\dd t$, with
\begin{equation}
 \mathcal D_T(t)=\frac{2F^{3\mathrm m}(t)}{vw}
 \left(\frac1t-\frac1{t+(1-w)/w}+\frac1{t-1}-\frac1{t+u/w}\right),
 \label{eq:three-point-triangle-density}
\end{equation}
where $F^{3\mathrm m}(t)=F^{3\mathrm m}(z(t)v,(1-z(t))w)$.
Both integrands can now be integrated in generalized polylogarithms
(GPLs)~\cite{Panzer}.
Let $A_B(z)$ and $A_T(t)$ be primitives of $\mathcal D_B(z)$ and
$\mathcal D_T(t)$, respectively, with vanishing constant terms in
their expansions at $z=0$ and $t=0$.

The two primitives use different coordinates, so their endpoint
constants must be combined using the original cutoff in $z$.
Equation~\eqref{eq:three-point-rationalization} gives
\begin{equation}
 t_- =\frac{u}{1-w}\delta+O(\delta^2),\qquad
 1-t_+=\frac{\delta}{1-v}+O(\delta^2).
 \label{eq:three-point-cutoff-map}
\end{equation}
To determine the resulting finite correction, expand the triangle
function at its two endpoints:
\begin{align*}
 F^{3\mathrm m}(t)&=\frac12\log w\,\log t+F_0^{3\mathrm m}
 +O(t\log t), &&t\to0,\\
 F^{3\mathrm m}(t)&=\frac12\log v\,\log(1-t)+F_1^{3\mathrm m}
 +O((1-t)\log(1-t)), &&t\to1.
\end{align*}
The finite coefficients are
\begin{align}
 F_0^{3\mathrm m}&=\Li_2(1-w)+\frac12\log\frac{v(1-w)}u\log w,
 \nonumber\\
 F_1^{3\mathrm m}&=\Li_2(1-v)+\frac12\log v\log[w(1-v)].
 \label{eq:three-point-triangle-endpoints}
\end{align}
Writing $\ell_0=\log[u/(1-w)]$ and $\ell_1=\log[1/(1-v)]$, integration
of these endpoint terms with the limits
\eqref{eq:three-point-cutoff-map} yields
\begin{equation}
 C_{\rm end}=
 -\frac{\tfrac12\log w\,\ell_0^2+2F_0^{3\mathrm m}\ell_0}{vw}
 +\frac{\tfrac12\log v\,\ell_1^2+2F_1^{3\mathrm m}\ell_1}{vw}.
 \label{eq:three-point-endpoint-correction}
\end{equation}
The divergent endpoint logarithms cancel between the two sectors, leaving
\begin{equation}
 K(u,v,w)=\Reg_{z=1}A_B(z)+\Reg_{t=1}A_T(t)+C_{\rm end}.
 \label{eq:three-point-endpoint-formula}
\end{equation}
Here $\Reg$ extracts the constant term in the endpoint expansion in
powers of $\log(1-z)$ or $\log(1-t)$.

\paragraph{Reconstruction of the remainder.}
We add the cyclic images of $K$ as in \eqref{eq:three-point-D} to form
$\omega_{\qb}$ in \eqref{eq:three-point-one-form}.  In the positive
coordinates \eqref{eq:positive-chart}, we integrate its $x$ component
from the collinear boundary $x=0$ at fixed $y$, with vanishing boundary
value, and denote the result by $R_{\qb}(x,y)$.
Using \texttt{HyperInt}~\cite{Panzer}, we reduce the source and the derivatives of
this primitive to a common GPL basis and verify
\begin{equation}
 \partial_xR_{\qb}=(\omega_{\qb})_x,\qquad
 \partial_yR_{\qb}=(\omega_{\qb})_y,\qquad
 R_{\qb}(0,y)=0.
 \label{eq:closedness-check}
\end{equation}
The $y$ identity is an independent compatibility check, since only the
$x$ component was integrated.  The reconstructed function agrees with
\eqref{eq:explicit-three-point-remainder}, including the transcendental
constants.

\section{Details of the MHV last-entry analysis}
\label{app:mhv-last-entries}

We present the collinear residue calculations underlying the MHV
last-entry candidates of section~\ref{sec:mhv-last-entries} and compare
the candidate space with other proposed last-entry structures.
All labels lie on the periodic cover, with the
monodromy and insertion conventions of appendix~\ref{app:monodromy}.

\subsection{Five-bracket contributions}
\label{app:last-entry-residues}

To exhibit a standard five-bracket and one multiplied by a rational
function of four-brackets, insert $\mathcal Z_\star$ between
$\mathcal Z_3$ and $P\mathcal Z_1$ on a three-point contour. Write
$\mathcal Z_a^\pm=P^{\pm1}\mathcal Z_a$, including the insertion and
its images. Two representatives in the resulting four-point source are
\begin{align}
 Y_1&=[2,3,\star,1^+,1],\nonumber\\
 Y_2&=
 \frac{\br{1,3^-,\star^-,\star}\br{1,3,\star,3^-}}
      {\br{1^+,3,\star,1}\br{\star,\star^-,3,3^-}}
 [3^-,\star^-,3,\star,1].
 \label{eq:last-entry-source-representatives}
\end{align}
The four- and five-brackets are defined in
\eqref{eq:nmhv-ordered-brackets}. These representatives occur in the box
coefficients \eqref{eq:nmhv-box-coefficient} and their cyclic images.
The second representative contains both $\star$ and $\star^-$, which
share the same Grassmann coordinate at $\gamma=0$.

Let $\dd\tau\,\cK_3[Y]$ denote the Grassmann integral and collinear
residue of a prefactor $Y$, including the dotted measure in
\eqref{eq:insertion}, before the longitudinal integration. We use
\begin{equation}
 Z_\star=Z_3-\varepsilon Z_2+c_3\varepsilon\tau Z_1^+
             +c_3'\varepsilon^2Z_2^+,
 \qquad c_3=\frac{\br{2,3}}{\br{3,1^+}}.
 \label{eq:last-entry-five-bracket-curve}
\end{equation}
The Grassmann integration is performed before restricting the remaining
$\chi_\star^A$ to the same curve. If the argument of the Grassmann
delta function is $a_\star\chi_\star+\psi$, with $\psi$ independent
of $\chi_\star$, the projection gives
\begin{equation}
 \int(\dd^3\chi_\star)^A\,
 \delta^{0|4}(a_\star\chi_\star+\psi)
 =a_\star^3(a_\star\chi_\star+\psi)^A.
 \label{eq:last-entry-five-bracket-projection}
\end{equation}

For $Y_1$, $a_\star=\br{2,3,1^+,1}$. After the restriction,
the argument of the delta function is $c_3'\varepsilon^2\Xi$, where $\Xi$
is the linear Grassmann argument of the delta function in
$[2,3,2^+,1^+,1]$. The five bosonic denominator factors together start
at order $\varepsilon^4$. Multiplying the projected expression by the
insertion measure and taking the residue gives
\begin{equation}
 \dd\tau\,\cK_3[Y_1]^A_{\dot\alpha}
 =-\frac{(2,3,1^+)_{\dot\alpha}\Xi^A}
 {\br{2,3,1^+,1}\br{2,3,1^+,2^+}}\,\frac{\dd\tau}{\tau}
 =\qb^A_{\dot\alpha}\log
 \frac{\br{2,3,1^+,2^+}}{\br{2,3,1^+,1}}\,\frac{\dd\tau}{\tau}.
 \label{eq:last-entry-five-bracket-rational}
\end{equation}
The second equality follows by differentiating the four-brackets with
$\chi_i^+=\chi_i$.
The transverse coefficient $c_3'$ has canceled.

For $Y_2$, one must first combine the coefficients of
$\chi_\star$ and $\chi_{\star^-}$ in the argument of the delta function:
\begin{equation}
 a_\star=\br{3,\star,1,3^-}+\br{1,3^-,\star^-,3}.
 \label{eq:last-entry-five-bracket-shared-fermion}
\end{equation}
We then apply \eqref{eq:last-entry-five-bracket-projection}, with
$\chi_{3^-}=\chi_3$, and include the four-bracket multiplier in $Y_2$
before taking the residue. This multiplier is generically of order
$\varepsilon^{-1}$, so its collinear behavior contributes to the result.
To express both residues as logarithmic kernels, set
$X=(Z_3,Z_2-c_3\tau Z_1^+)$, with
$\br{X,a,b}=\det(Z_3,Z_2-c_3\tau Z_1^+,Z_a,Z_b)$.
The longitudinal factors obtained from $Y_1,Y_2$ are, respectively,
\begin{equation}
 \frac1\tau=\partial_\tau\log\frac{\br{X,1,2}}{\br{X,1^+,2^+}},
 \qquad
 -\frac{c_3\br{3,1^+,1,3^-}}{\br{X,1,3^-}}
 =\partial_\tau\log\br{X,1,3^-}.
 \label{eq:last-entry-five-bracket-longitudinal}
\end{equation}
Thus the projected expressions take the form
\begin{align}
 \dd\tau\,\cK_3[Y_1]
 &=\dd_\tau\log\frac{\br{X,1,2}}{\br{X,1^+,2^+}}\,
   \qb\log\frac{\br{2,3,1^+,2^+}}{\br{2,3,1^+,1}},\nonumber\\
 \dd\tau\,\cK_3[Y_2]
 &=\dd_\tau\log\br{X,1,3^-}\,
   \qb\log\frac{\br{2,3,1^+,3^+}}{\br{2,3,1^+,1}}.
 \label{eq:last-entry-local-residues}
\end{align}
Here $\dd_\tau\log f=(\partial_\tau\log f)\dd\tau$, and the free
supersymmetry and dotted indices are suppressed. In each line, the
$\dd_\tau\log$ term gives the longitudinal integration kernel, while
the $\qb\log$ term depends only on the original momentum twistors.
After multiplying by the coefficient function of $Y$ in the source,
the $\qb\log$ term can therefore be taken outside the $\tau$ integral,
with the common endpoint prescription understood.

The same procedure applies at arbitrary multiplicity: the Grassmann
integration and collinear residue are evaluated using the same insertion
parametrization, with all periodic images sharing the insertion's
Grassmann coordinate. The four-point examples above illustrate the
essential steps.

\subsection{Triangle contributions}
\label{app:last-entry-triangles}

We extract the last-entry factors of the triangle coefficient
\eqref{eq:nmhv-triangle-coefficients} by using one cut solution as the
integration variable and decomposing the resulting rational kernel into
simple fractions. Consider a triangle with one corner on the insertion
edge and another at $x_s$, represented by $(Z_{s-1},Z_s)$ on the periodic
cover. The collinear insertion \eqref{eq:collinear-curve}, with
$c_e=\br{e-1,e}/\br{e,e+1}$, gives the leading dual coordinates
associated with $(Z_e,Z_\star)$,
\begin{equation}
 x_\star=x_e+\frac{\tau}{1+\tau}p_e,\qquad
 \theta_\star=\frac{\theta_e+\tau\theta_{e+1}}{1+\tau}.
 \label{eq:last-entry-triangle-coordinates}
\end{equation}
We take $h=h_+(\tau)$ as the integration variable. Since $p_e^2=0$,
the cut condition $[hq-(x_\star-x_s)]^2=0$ gives
\begin{equation}
 \begin{gathered}
 \tau=\frac{N(h)}{J(h)-N(h)},\qquad
 N(h)=\bigl[hq-(x_e-x_s)\bigr]^2,\\
 J(h)=2p_e\cdot\bigl[hq-(x_e-x_s)\bigr],\qquad
 \lambda_h^\alpha=-\bigl[hq-(x_e-x_s)\bigr]^{\alpha\dot\beta}
                       \widetilde\lambda_{e\dot\beta}.
 \end{gathered}
 \label{eq:last-entry-triangle-conic}
\end{equation}
The last expression fixes a rational normalization of the cut spinor:
the term proportional to $p_e$ drops out when the cut momentum acts on
$\widetilde\lambda_e$. Writing $y=h_+-h_-$, the Jacobian is
$\dd\tau/[(1+\tau)^2q^2y]=\dd h/J(h)$.
The Grassmann integral and collinear residue for this cut solution,
including the factor $-2$ in the one-loop input and the dotted measure,
then give
\begin{equation}
 \Omega_{\dot\alpha}(h)=
 2\br{s-1,s}\widetilde\lambda_{e\dot\alpha}
 \frac{(\theta_s-\theta_\star)^\beta\lambda_{h\beta}}
 {h(1-h)\br{s-1,h}\br{s,h}}\,\dd h ,
 \label{eq:last-entry-triangle-single-root}
\end{equation}
where the supersymmetry index is suppressed.

The $\qb\log$ factors are the coefficients in the partial fraction
decomposition of $\Omega(h)$. With the cut spinor normalization in
\eqref{eq:last-entry-triangle-conic}, the spinor brackets satisfy
\[
 \br{j,h}=-\frac{(1-h)\br{e-1,e,e+1,j}
                  +h\br{e-1,e,e+1,j^+}}
 {\br{e-1,e}\br{e,e+1}},\qquad j=s-1,s,
\]
where $Z_j^+=PZ_j$. The common denominator depends only on the external
$\lambda$ spinors and is annihilated by $\qb$.
Since $\theta_{e+1}-\theta_e$ is proportional to $\lambda_e$ and
$\br{e,h}=J(h)$, the apparent denominator $J(h)$ in the numerator
of $\Omega(h)$ cancels.
For generic kinematics, the four simple
poles are $h=0,1$ and $h=h_{s-1},h_s$, where $h_j$ is the zero of
$\br{j,h}$ for $j=s-1,s$. Direct evaluation gives
\begin{equation}
 \begin{alignedat}{2}
 \Res_{h=0}\Omega(h)&=2\qb\log L_0,
 &\qquad L_0&=\frac{\br{e-1,e,e+1,s}}{\br{e-1,e,e+1,s-1}},\\
 \Res_{h=1}\Omega(h)&=2\qb\log L_1,
 &\qquad L_1&=\frac{\br{e-1,e,e+1,(s-1)^+}}{\br{e-1,e,e+1,s^+}},\\
 \Res_{h=h_{s-1}}\Omega(h)&=2\qb\log L_2,
 &\qquad L_2&=\frac{\br{e-1,e,e+1,s-1}}{\br{e-1,e,e+1,(s-1)^+}},\\
 \Res_{h=h_s}\Omega(h)&=2\qb\log L_3,
 &\qquad L_3&=\frac{\br{e-1,e,e+1,s^+}}{\br{e-1,e,e+1,s}}.
 \end{alignedat}
 \label{eq:last-entry-triangle-residues}
\end{equation}
Since $\Omega(h)=\mathcal O(h^{-2})\dd h$ at
infinity, these residues determine the differential.
At the same insertion point, the other cut solution is
$\bar h=2q\cdot(x_\star-x_s)/q^2-h$, with
$\dd\bar h/J(\bar h)=-\dd h/J(h)$.
The physical coefficient adds both cut contributions with the same
$1/(q^2y)$, so this Jacobian gives
\begin{equation}
 \begin{aligned}
 \Omega(h)-\Omega(\bar h)
 ={}&2(\qb\log L_0)\,\dd_h\log\frac{h}{\bar h}
 +2(\qb\log L_1)\,\dd_h\log\frac{h-1}{\bar h-1}\\
 &+2(\qb\log L_2)\,\dd_h\log\frac{h-h_{s-1}}{\bar h-h_{s-1}}
 +2(\qb\log L_3)\,\dd_h\log\frac{h-h_s}{\bar h-h_s},
 \end{aligned}
 \label{eq:last-entry-triangle-orbit}
\end{equation}
where $\Omega(\bar h)$ includes $\dd\bar h$.
Each $L_r$ is rational in the external twistors and independent of $h$.
Both cut contributions multiply the same triangle function
$F^{3\mathrm m}(U,V)$, with $U=(x_\star-x_s)^2/q^2$ and
$V=[q-(x_\star-x_s)]^2/q^2$. The $\qb\log L_r$ factors can therefore
be taken outside the longitudinal integral, with the triangle function
remaining inside.

When the insertion and the adjacent external leg belong to the same
massive corner, retain $+\varepsilon^2\mathcal Z_{e+2}$ in the
supertwistor insertion. For the triangle corners $(Z_{e-1},Z_e)$ and
$(Z_\star,Z_{e+1})$, taking the branch with $h_-\to0$, the Grassmann
integral and collinear residue give
\begin{equation}
 2\qb\log\frac{\br{e-1,e,e+1,e^+}}
 {\br{e-1,e,e+1,e+2}}\,\frac{\dd\tau}{\tau}.
 \label{eq:last-entry-triangle-degenerate}
\end{equation}
This ratio has the same form as the last-entry candidates of
section~\ref{sec:mhv-last-entries} and is independent of $\tau$.
The $\qb\log$ factor can again be taken outside the regulated longitudinal
integral, with the triangle function remaining inside.

\subsection{Comparison of last-entry spaces}
\label{app:last-entry-space}

We compare our candidate space with the parity-even and parity-odd
candidate structures of ref.~\cite{Li:2024rkq}. The even generators are
the nonzero ratios
\begin{equation}
 \frac{\mathcal L_{j,i+n}\mathcal L_{i+n,j}}
      {\mathcal L_{ij}\mathcal L_{ji}},\qquad
 \frac{\mathcal L_{i,i+n}\mathcal L_{j,j+n}}
      {\mathcal L_{ij}\mathcal L_{ji}},
 \qquad 2\leq j-i\leq n-1,
 \label{eq:last-entry-neutral-ratios}
\end{equation}
and their cyclic images, with $\mathcal L_{ij}$ defined in
\eqref{eq:last-entry-periodic-brackets}. The odd generators are
\begin{equation}
 \frac{\br{i,i+1}}{[i,i+1]},\qquad
 \frac{\sum_{k=i}^{j}\br{i,k}[k,j]}
      {\sum_{k=i}^{j}[i,k]\br{k,j}},
 \qquad 2\leq j-i\leq n-1,
\end{equation}
where $[i,j]$ denotes the square spinor bracket. Their logarithmic
combinations must have zero total little-group weight. Factoring the rational
arguments in generic momentum twistors and comparing their exponent
vectors over $\mathbb Q$ gives
\begin{center}
\begin{tabular}{c c c c}
\toprule
$n$ & parity even & parity odd & combined dimension\\
\midrule
3 & 3 & 0 & 3\\
4 & 8 & 4 & 12\\
5 & 15 & 10 & 25\\
6 & 24 & 18 & 42\\
\bottomrule
\end{tabular}
\end{center}
For $n=3,4,5,6$, the parity-even and parity-odd structures together span
the same space as our last-entry candidates.

\section{Derivation of the Mellin identity}
\label{app:fermion-mellin}

We derive \eqref{eq:fermion-dispersion-test} for the three-point source.
Throughout this appendix,
$u,v,w>0$, $u+v+w=1$, $s=v+w$, and logarithms are real for $\tau>0$.

The substitution $z=\tau/(1+\tau)$ reduces the one-loop primitive
\eqref{eq:three-point-log-primitive} to
\begin{equation}
 \mathcal P_\ell\!\left(\frac{\tau}{1+\tau}\right)
 =\frac{L_u(\tau)}{vs}+\frac{L_{u^{-1}}(\tau)}{ws},
 \qquad
 L_a(\tau)=\log(1+\tau/a)\log\frac{\tau}{\tau+a}.
 \label{eq:log-primitive-tau}
\end{equation}
To transform it, write
\begin{equation}
 \begin{split}
 b_a(\tau)&=\frac1{\tau+a}-\frac1\tau,\qquad
 g_a(\tau)=2\partial_\tau L_a(\tau),\\
 k_T&=\frac{b_u}{vs}+\frac{b_{u^{-1}}}{ws},\qquad
 k_\ell=\frac{g_u}{vs}+\frac{g_{u^{-1}}}{ws}.
 \end{split}
 \label{eq:mellin-elementary-densities}
\end{equation}
For $a>0$ and $0<\operatorname{Re}\nu<1$, a convergent beta integral gives
\begin{equation}
 \int_0^\infty\dd\tau\,\tau^\nu b_a(\tau)
 =-a^\nu\Gamma(\nu)\Gamma(1-\nu)
 =-\frac{\pi a^\nu}{\sin\pi\nu}.
 \label{eq:mellin-tree-beta}
\end{equation}
For the logarithmic density, set $t=\tau/a$ to obtain
\begin{equation}
 g_a(\tau)=\frac2a\left[
 \frac{\log t}{1+t}+\frac{1-t}{t(1+t)}\log(1+t)\right].
 \label{eq:mellin-log-elementary}
\end{equation}
Its endpoint behavior is
\begin{equation}
 \begin{aligned}
 g_a(\tau)&=\frac2a[\log(\tau/a)+1]
       +O(\tau\log\tau),&&\tau\to0,\\
 g_a(\tau)&=\frac{2a}{\tau^2}[\log(\tau/a)-1]
       +O(\tau^{-3}\log\tau),&&\tau\to\infty.
 \end{aligned}
 \label{eq:mellin-log-endpoints}
\end{equation}
Its Mellin transform is therefore holomorphic in
$-1<\operatorname{Re}\nu<1$. To evaluate the separate terms in
\eqref{eq:mellin-log-elementary}, first take
$-1<\operatorname{Re}\nu<0$, where each integral converges. With
$\mathcal B(\nu)=\Gamma(1+\nu)\Gamma(-\nu)=-\pi/\sin\pi\nu$,
beta differentiation and an integration by parts give
\begin{align}
 \mathcal A(\nu)&=\int_0^\infty\dd t\,
 \frac{t^\nu\log t}{1+t}
 =\mathcal B(\nu)[\psi(1+\nu)-\psi(-\nu)],\nonumber\\
 \mathcal C(\nu)&=\int_0^\infty\dd t\,t^{\nu-1}\log(1+t)
 =-\frac{\mathcal B(\nu)}\nu,\nonumber\\
 \mathcal D(\nu)&=\int_0^\infty\dd t\,
 \frac{t^\nu\log(1+t)}{1+t}
 =\mathcal B(\nu)[\psi(1)-\psi(-\nu)].
 \label{eq:mellin-beta-derivatives}
\end{align}
It follows that
\begin{equation}
 \int_0^\infty\dd\tau\,\tau^\nu g_a(\tau)
 =2a^\nu(\mathcal A+\mathcal C-2\mathcal D)
 =a^\nu\mathcal B(\nu)h_\psi(\nu),
 \label{eq:mellin-log-beta}
\end{equation}
where the last equality uses $\psi(1-\nu)=\psi(-\nu)-1/\nu$.
The apparent pole at $\nu=0$ on the right is cancelled by the double
zero of $h_\psi$. Both sides thus continue holomorphically to the full
logarithmic convergence strip. Combining
\eqref{eq:mellin-elementary-densities}, \eqref{eq:mellin-tree-beta}
and \eqref{eq:mellin-log-beta} proves
\eqref{eq:fermion-dispersion-test} in the tree convergence strip,
without assigning values to separately divergent integrals.

\bibliographystyle{JHEP}
\bibliography{references}

\end{document}